\documentclass[12pt,onecolumn,oneside,a4paper,peerreview]{IEEEtran}
\pdfoutput = 1
\usepackage{cite}
\usepackage{graphicx}
\usepackage{amsmath}
\usepackage{times}
\usepackage{latexsym}
\usepackage{graphicx}
\usepackage{bm}
\usepackage{amssymb}
\usepackage[center]{caption2}
\usepackage{stfloats}
\usepackage{cases}
\usepackage{url}
\usepackage{array}
\usepackage{setspace}
\usepackage{fancyhdr}
\usepackage{color}
\usepackage{subfigure}
\usepackage{bbding}
\usepackage{hyperref}
\hypersetup{hypertex=true,
	colorlinks=true,
	linkcolor=blue,
	anchorcolor=blue,
	citecolor=blue}

\newtheorem{theorem}{\bf{Theorem}}
\newtheorem{lemma}{\bf{Lemma}}
\newtheorem{corollary}{\bf{ Corollary}}
\newtheorem{proposition}{\bf {Proposition}}

\def\proof{\noindent\hspace{2em}{\itshape Proof: }}
\def\endproof{\hspace*{\fill}~$\square$\par\endtrivlist\unskip}

\begin{document}
\title{Over-the-Air Federated Edge Learning with Error-Feedback One-Bit Quantization and Power Control}

\author{Yuding Liu, Dongzhu Liu, Guangxu Zhu, Qingjiang Shi and Caijun Zhong}

\maketitle

\begin{abstract}
	\emph{Over-the-air federated edge learning} (Air-FEEL) is a communication-efficient framework for distributed machine learning using training data distributed at edge devices.{ This framework enables all edge devices to transmit model updates simultaneously over the entire available bandwidth, allowing for over-the-air aggregation.} A \emph{one-bit digital over-the-air aggregation} (OBDA) scheme has been recently proposed, featuring one-bit gradient quantization at edge devices and majority-voting based decoding at edge server. However, the low-resolution one-bit gradient quantization slows down the model convergence and leads to performance degradation. On the other hand, the aggregation errors caused by channel fading in Air-FEEL is remained to be solved. {To address these issues, we propose the \emph{error-feedback one-bit broadband digital aggregation} (EFOBDA) and an optimized power control policy. To this end, we first provide a theoretical analysis to evaluate the impact of error feedback on the convergence of Air-FEEL with EFOBDA. The analytical results show that, by setting an appropriate feedback strength, EFOBDA is comparable to the Air-FEEL without quantization, thus enhancing the performance of OBDA. Then, we further introduce a power control policy by maximizing the convergence rate under instantaneous power constraints. The convergence analysis and optimized power control policy are verified by the experiments, which show that the proposed scheme achieves significantly faster convergence and higher test accuracy in image classification tasks compared with the one-bit quantization scheme without error feedback or optimized power control policy.}
\vspace{1cm}
\begin{center}
	{\bf Index Terms}
\end{center}
Over-the-air, Federated learning, Error-feedback, Power control.
\end{abstract}

\clearpage

\section{Introduction}
Recent breakthroughs in artificial intelligence (AI) motivate the development of AI technologies at the network edge\cite{zhu2019broadband}. A wealth of data generated by edge devices such as smart mobile phones has injected vitality into \emph{edge learning} but has also raised concerns about data privacy\cite{park2019wireless}. On this account, \emph{federated edge learning} (FEEL) has been proposed to distribute the model training task over edge devices by using distributed local data without compromising their privacy\cite{mcmahan2017communication,yang2019federated,samarakoon2019distributed}. Generally, the FEEL framework implements the \emph{stochastic gradient descent} in a distributed manner. Gradient updates computed by edge devices are transmitted to the edge server and aggregated to update a global model. Edge devices involved in model training communicate with the edge server through a \emph{multiple access channel} (MAC), which leads to a communication bottleneck due to the high dimension of gradient updates. To cope with this issue, a communication-efficient FEEL framework called over-the-air FEEL (Air-FEEL) has been proposed in \cite{zhu2021over}. By exploiting the property of waveform superposition in non-orthogonal MAC, Air-FEEL allows all edge devices to upload gradient updates simultaneously, which are aggregated over the air. {Compared to traditional digital orthogonal multiple access methods that separate communication and computation, Air-FEEL achieves computation, e.g., the aggregation of local updates, via communication, which accelerates the learning speed and enhances the communication efficiency.}

The idea of \emph{over-the-air computation} (AirComp) was proposed for data aggregation in the study of sensor networks to cope with channel distortion introduced by MAC  \cite{nazer2007computation}. Researchers considered the transmission of linear functions of two correlated Gaussian sources in a distributed manner and proposed a lattice coding scheme which was shown to be better in performance than uncoded transmission \cite{soundararajan2012communicating}. The high bandwidth efficiency of analog AirComp attracts lots of attention which drives more studies in this area \cite{wang2011distortion,goldenbaum2014channel,goldenbaum2015achievable,goldenbaum2013robust,abari2015airshare,zhu2018mimo,li2019wirelessly,wen2019reduced}. Several practical implementations of AirComp were designed with a synchronization system over sensors \cite{goldenbaum2013robust,abari2015airshare}. {To enable high dimensional function computation, AirComp has been implemented in \emph{multiple-input-multiple-output} (MIMO) channels \cite{zhu2018mimo}, which was then extended to wireless-powered AirComp system \cite{li2019wirelessly} and massive MIMO AirComp system \cite{wen2019reduced}.}

Recently, implementing AirComp in FEEL has attracted much attention due to the advantages in transmitting an aggregation of high-dimensional updates \cite{sery2020analog}. The study of the Air-FEEL system mainly focuses on several research directions: learning rate optimization \cite{xu2021learning}, device scheduling \cite{sun2021dynamic,xia2021fast,fan2021jointa}, gradient compression \cite{amiri2020machine,amiri2020federated,fan20211} and power control\cite{cao2021optimized,yang2022over,zhang2021gradient,guo2022joint}. For instance, to mitigate the wireless distortion, {the authors in \cite{xu2021learning} adapt local learning rate to the time-varying channels}. An energy-aware dynamic device scheduling algorithm is designed in\cite{sun2021dynamic} to optimize the training performance of FEEL under a total energy consumption constraints of devices. The threshold-based device selection scheme proposed in \cite{xia2021fast} aims to achieve reliable uploading of local models. {The authors in \cite{fan2021jointa} develop a joint optimization scheme for accurate FL implementation}, which allows the parameter server to select a subset of workers and determine an optimized power scaling factor. As for gradient compression, the authors in \cite{amiri2020machine,amiri2020federated} propose a source-coding algorithm exploiting gradient scarification, and a compressive-sensing-based gradient aggregation approach is developed in \cite{fan20211} to further improve the communication efficiency. To deal with the aggregation error caused by the channel fading and noise perturbation, a transmission power control policy is needed \cite{cao2021optimized}. In the direction that focuses on power control, prior works \cite{yang2022over,zhang2021gradient,guo2022joint} have considered channel inversion and its variants and minimization of the individual aggregation distortion.

Lately, the authors in \cite{zhu2020one} proposed an Air-FEEL framework based on digital modulation, called OBDA, which features one-bit quantization and modulation at the edge devices and majority-vote as a decoder at the edge server. However, one-bit quantization changes the direction of the gradient descent step, which slows down the convergence and leads to performance degradation. As for the power control in \cite{zhu2020one}, \emph{truncated channel-inversion} was adopted to align the channel gains among the selected active devices, which, however, may lead to error in estimating the global gradient due to the information loss of the truncated local gradients. \cite{karimireddy2019error} addressed the convergence issue of SGD with one-bit quantization by incorporating the error feedback mechanism. However, it considers centralized SGD and the implementation of error feedback enabled SGD with gradient quantization in Air-FEEL remains unexplored, which thus motivates the current work.

This paper studies a FEEL system consisting of multiple edge devices and one edge server. Inspired by the centralized error-feedback algorithm in \cite{karimireddy2019error}, we consider distributed SGD over non-orthogonal MAC and propose the one-bit digital aggregation with error-feedback (EFOBDA) to enable gradients aggregation over-the-air. Then, we further optimize the power control parameters at each communication round. One of the main contributions of this work is the analytical study of the convergence behavior of FEEL with error-feedback in the wireless setting. The main contributions are summarized as follows.

\begin{itemize}
	\item {\bf Convergence analysis:}
	The convergence results of EFOBDA are derived for two scenarios: 1) general fading channels and 2) Gaussian channels. The convergence results are comprised by initialization, aggregation error incurred by wireless transmissions, and the error incurred by quantization and stochastic gradient, with a scaling factor in terms of communication round and error-feedback strength. Increasing error-feedback strength can reduce the quantization error but at a cost of convergence speed. In the scenario of AWGN channels, by setting a time decaying learning rate the error mentioned above can reduce to zero as iteration going on. However, in the scenario of fading channel, the additional signal misalignment error becomes convergence bottleneck, which can be mitigated by the power control policy, and thus motivate the optimization problem discussed in the next.
	
	\item {\bf Power control optimization:}
	Given the convergence analysis, we optimize the convergence rate over the power control parameters under the transmit power constraint. The problem is equivalent as minimizing the aggregation error and can be addressed in parallel of $T$ communication rounds. For each round, the problem is shown to be convex and can be solved with a closed-form solution.
	
	\item {\bf Performance evaluation:}
	Extensive experiments on the MNIST dataset and the CIFAR-10 dataset are conducted to demonstrate the effectiveness of the proposed method. It is shown that the proposed scheme achieves significantly faster convergence than {the one-bit quantization scheme without error-feedback and optimized power control policy}, {as the error-feedback algorithm and the proposed power control policies can better handle the gradient information loss induced by quantization and aggregation errors. Besides, the convergence rate of the proposed scheme is comparable with the schemes without gradient quantization.}
	
\end{itemize}
\emph{Organization}: The remainder of the paper is organized as follows. Section II introduces the learning and communication models of proposed EFOBDA scheme. Section III presents the convergence analysis under different channel models. Section IV presents the formulated power control optimization problem and corresponding optimal solutions. Section V presents the experimental results using real datasets followed by concluding remarks in Section VI.

\section{System Model} \label{sec:system model}
\begin{figure}[t]
	\centering
	\includegraphics[width=14cm]{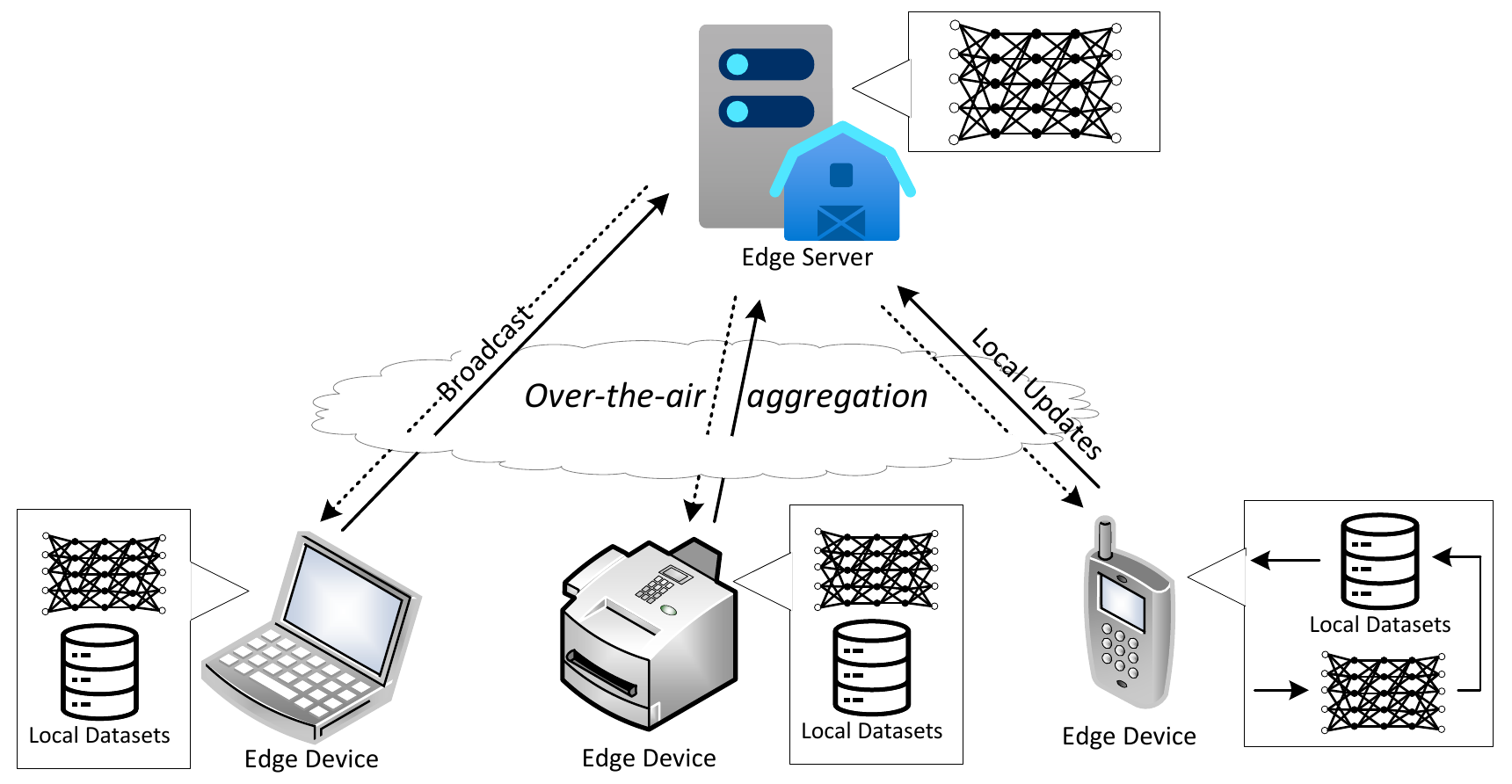}
	\caption{The architecture of Air-FEEL.}
	\label{system_model}
	\vspace{-4mm}
\end{figure}

We consider a federated edge learning (FEEL) system consisting of an edge server and $K$ devices as shown in Fig.\ref{system_model}. Each edge device $ k $ has its own local data-set $ {\cal D}_k $ encompassing $ D_k $ pairs of data samples $ {\mathcal D}_k=\{{(s_i, l_i)}\}_{i=1}^{D_k} $, where $ s_i $ is the feature vector and $ l_i $ is the label. For simplicity, we consider each device has the identical number of local samples, i.e., $ D_k=D $ for all $ K $ devices. The generalization to heterogeneous local data-set sizes is straightforward by adding a scaling factor.

\subsection{Learning Model}
We denote $f({\bf w},{\bf s}_i, l_i)$ as the loss function on data sample $({\bf s}_i, l_i)$ with {model vector} ${\bf w} \in \mathbb{R}^q$. The local loss function on ${\cal D}_k$ is
\begin{align}\label{eq:local_loss} 
F_k({\bf w}) = \frac{1}{D} \sum_{({\bf s}_i, l_i) \in {\cal D}_k} f({\bf w},{\bf s}_i, l_i).
\end{align}
Then, the global loss function for all devices evaluated at {model vector} ${\bf w}$ is given by:
\begin{align}\label{eq:global_loss} 
F({\bf w}) =\frac{1}{K} \sum_{k = 1}^K F_k({\bf w}).
\end{align}

The training process aims to find a model vector ${\bf w}$ by minimizing the global loss function $F({\bf w})$ as
\begin{align}\label{eq:goal} 
{\bf w}^* = \arg \min F({\bf w}).
\end{align}

The learning protocol coordinates iterations between edge devices and the server as detailed below. At each communication round $ t $, the server broadcasts the global model $ {\bf w}^{(t)} $ to the devices. Upon receiving the model, each device randomly {samples a  mini-batch of training examples as $\hat{\cal D}_k $} and computes the local gradient:
\begin{align}\label{eq:local_update}
{\bf g}_{k}^{(t)} = \frac{1}{n_b} \sum_{({\bf s}_i, l_i) \in \hat{\cal D}_k} \nabla f ({\bf w}^{(t)}, {\bf s}_i, l_i),
\end{align}
which is shagreen with the edge server. The edge server aggregates the local gradient and updates the global model as follows
\begin{align}\label{eq:gradient_averaging}
	{\bf w}^{(t+1)} = {\bf w}^{(t)}-\eta \frac{1}{K} \sum_{k=1}^K {\bf g}_{k}^{(t)},
\end{align}
where $ \eta $ is the learning rate. The steps (4) and (5) iterate until the number of communication rounds reaches the communication overhead budget or a convergence condition is met. 

\subsection{Communication Model}

In the upload stage, each edge device transmits the local gradient to the edge server over a shared multiple access channel. It is observed from \eqref{eq:gradient_averaging} that only the aggregated gradient is needed at the edge server. This motivates the communication-efficient scheme -- AirComp, which exploits the superposition property of wireless channels. In this setup, all the devices simultaneously transmit their local gradients which are aggregated over the air.

We employ the OFDM modulation for communication to deal with the inter-symbol interference and frequency selective channel fading. The whole bandwidth $ B $ is divided into $ M $ orthogonal sub-channels, and thus,  we use $ q/M $ OFDM symbols to transmit a gradient vector. For simplicity, we assume that the channel coefficients remain unchanged within each communication round, but may change over different rounds. At each communication round $ t $, the device $k$ transmits the modulated symbol ${x}_k^{(t)}[i]$ over the $ m $-th sub-channel. The received aggregated symbol at the server is given by
\begin{align}\label{over_the_air_aggre} (\text{Over-the-air aggregation}) \quad
	{y}^{(t)}[i] = \sum_{k =1}^K {{ h}_{k}^{(t)}[i] { p}_{k}^{(t)}[i]  {x}_k^{(t)}[i]} + {z}[i],
\end{align}
where $ h_{k}^{(t)} [i] $ is the channel coefficient and  $p_{k}^{(t)} [i]$ is the transmission power. As for the channel noise $ {z}[i] $, we consider \textit{additive white Gaussian noise} (AWGN), i.e., ${{z}[i]} \sim \mathcal{CN}\left( 0,\sigma_z^2\right)$ with the noise power $ \sigma_z^2 $.

Furthermore, we have the transmit power constraint on each device over the whole training period. We consider the normalized transmitted signal $ {\mathbb{E}}\left[|{x}_k^{(t)}[i]|^2\right]=1 $. Under the assumption of identical distribution of each sub-channels, the power constraint is given by 
\begin{align} \label{power_con}
	{ |{p}_{k}^{(t)}[i]|}^2  \leq \frac{P_0}{M}, \forall k,t.
\end{align}

\subsection{Basic Assumptions}\label{sec:assumptions}

We have the following assumptions on the compression scheme, loss function and gradient. 

{\bf Assumption 1} (\emph{Strictly} $(1-\delta)$-\emph{contractive Operator} [32]) We assume that the compression operator ${\mathcal C}$: ${\mathbb R}^q \rightarrow {\mathbb R}^q$ has the $(1-\delta)$-strictly contractive property with $\delta\in[0,1]$, i.e., 
\begin{align}
	\|{\mathcal C}({\bf a})-{\bf a}\|_2^2 \leq (1-\delta)\|{\bf a}\|_2^2,\quad \forall {\bf a} \in {\mathbb R}^q. \notag
\end{align}
Examples of $(1-\delta)$-contractive operator include: i) the sign operator as implemented in the later section, ii) top-$k$ which selects $k$ coordinates in $\bf a$ with the largest absolute value while zeroing out the rest [33].

{\bf Assumption 2} (\emph{Bounded Loss Function}) Given a model vector $ \bf w $, the global loss function is lower bounded by a value $ F^* $, i.e., $F({\bf w}) \geq F^*$, $\forall {\bf w}$.

{\bf Assumption 3} (\emph{Lipschitz Continuous Gradient}) The global loss function $ F({\bf w}) $ is differentiable and the gradient $ \nabla F ({\bf w}) $ is Lipschitz continuous with constant $ L $, i.e., 
\begin{align}\label{eq:smoothness}
	\!\!\! |F({\bf w}') - [F({\bf w}) + {\nabla F({\bf w})}^T({\bf w}' - {\bf w})]| \leq \frac{L}{2} \|{\bf w}' - {\bf w}\|^2, \qquad \forall {\bf w}', {\bf w} \in \mathbb{R}^q . \notag
\end{align}

{\bf Assumption 4} (\emph{Statistics of Gradient}) It is assumed that the {stochastic gradient} $\{{\bf g}_k^{(t)}\}_{k=1}^K$ defined in \eqref{eq:local_update} are independent and unbiased estimates of the global gradient ${\bf g}^{(t)} = \nabla F(\bf w)$ with  coordinate bounded variance, i.e.,
\begin{align}
	&\mathbb{E}[{\bf g}_k^{(t)}] = {\bf g}^{(t)}, \qquad \forall k,t, \notag \\ 
	&\mathbb{E}[( g_{k}^{(t)}[i]  -  g^{(t)}[i] )^2] \leq \sigma_i^2, \qquad \forall k, i, t, \notag \\
	&{\mathbb E}[||{\bf g}_{k}^{(t)}||^2] \leq G^2, \qquad \forall k,t, \notag
\end{align}
where $ g^{(t)}[i] $ is the $ i $-th element of $ {\bf g}^{(t)} $ and $\boldsymbol \sigma = [\sigma_1, \ldots, \sigma_q]$ is a vector of non-negative constants. 

\section{ One-Bit Broadband Digital Aggregation With Error-Feedback }\label{sec:EFOBDA}

In this section, we introduce the one-bit broadband digital aggregation with error-feedback (EFOBDA) for SignSGD. The aggressive quantization by using only one bit degrades the convergence speed, and thus motivates the use of error feedback to mitigate the impact of quantization error. To implement it in the wireless system, we first present the signal design at the transmitter which incorporates the error-feedback, gradient quantization, and digital modulation, and then introduce the designed post-processing at the receiver. In the last part of this section, we analyze the convergence behavior of the proposed FEEL framework. 

\subsection{Transceiver Design}

Inspired by the signSGD with error-feedback [30], we apply the error correction and one-bit quantization for the local gradient. In each communication round $ t $, each device computes the {local gradient} $ {\bf g}_{k}^{(t)} $ upon the current model  $ {\bf w}^{(t)} $ and then adds the quantization error $ {\bf e}_k^{(t)} $ to the {local gradient} as
\begin{align}(\text{Error Correction}) \quad 
	{\bf u}_{k}^{(t)} = \frac{1}{\beta} {\bf g}_{k}^{(t)} +  {\bf e}_{k}^{(t)},
\end{align} 
where $\beta$ controls the error-feedback strength. The {quantization error} $ {\bf e}_k^{(t)} $ compensates for the loss incurred by quantization.  We quantize the error corrected gradient ${\bf u}_{k}^{(t)}$ element-wisely as follows
\begin{align}(\text{One-bit Quantization}) \quad \label{quantization}
	\tilde{\bf u}_{k}^{(t)}= {\sf sign}({\bf u}_{k}^{(t)}).
\end{align}
The quantization error term ${\bf e}_{k}^{(t)}$ is then updated as follows
\begin{align} \label{accumulated_quantization_error}
	{\bf e}_{k}^{(t+1)} &= {\bf u}_{k}^{(t)} -  \tilde{\bf u}_{k}^{(t)}.
\end{align}
The quantization error is stored locally at device $ k $ and will be added to the local gradient in the next communication round. 

We consider \eqref{quantization} as a direct implementation of BPSK modulation on each element of $ {\bf u}_k^{(t)} $, where we have the modulated symbol in \eqref{over_the_air_aggre} as 
\begin{align}(\text{BPSK Symbol})
\quad x_k^{(t)}[i]= {\sf sign}\left({{u}_k^{(t)}}[i]\right).
\end{align}
The extension to higher order modulation schemes, e.g., QAM, is tractable by treating each modulated symbol as multiple orthogonal BPSK symbols. 

As per \eqref{over_the_air_aggre}, the server receives the modulated gradient elements in an aggregated form under the distortion of fading channel. For each communication round, the server cascades the received signals from multiple sub-channels and takes an average to decode the estimate of the global gradient as $ {\hat {\bf y}}^{(t)} = {{\bf y}}^{(t)}/K $, which is used for the local model update as 
\begin{align}\label{eq:new_model_update}
	{\bf w}^{(t+1)}={\bf w}^{(t)} -\eta {\hat {\bf y}}^{(t)} .
\end{align}

\subsection{Gradient Error Analysis}

We first consider the gradient error as the basis to develop the convergence analysis. At each communication round $ t $, the gradient error is defined as
\begin{align} \label{gradient_error}
	{\boldsymbol \xi}^{(t)} &= \sum_{k=1}^K{{\bf u}_{k}^{(t)}} -  {\bf y}^{(t)}.
\end{align}
In the following, we introduce the quantized gradient ${{\tilde{\bf u}}_k^{(t)}}$  to decompose the gradient error into two parts: quantization error and aggregation error as shown below.  
\begin{align}\label{relation}
	{\boldsymbol \xi}^{(t)} &= \underbrace{\sum_{k=1}^K{\left({\bf u}_{k}^{(t)} - \tilde{\bf u}_{k}^{(t)} \right)}}_{\sf Quantization \; error, \; {\bf e}_k^{(t+1)}}  + \underbrace{ \sum_{k=1}^K{\tilde{\bf u}_{k}^{(t)}} -  {\bf y}^{(t)}}_{\sf Aggregation \; error} \notag 
	\\ &= \sum_{k=1}^K \sum_{\tau=1}^t { \left( \frac{1}{\beta}{\bf g}_{k}^{(\tau)} - {\sf sign}({\bf u}_{k}^{(\tau)}) \right)} + \underbrace{\sum_{k =1}^K \left( 1 - {h}_{k}^{(t)} {p}_{k}^{(t)} \right) {\sf sign}\left({{\bf u}_k^{(t)}}\right)}_{\sf Signal \; misalignment \; error, \; {\bm \varepsilon}^{(t)}} + \bf z,
\end{align}
where the second equality is obtained by substituting (8) and \eqref{quantization} to \eqref{accumulated_quantization_error}. Equation \eqref{relation} follows that $ {\bf e}_{k}^{(0)} = 0 $. The aggregation error is comprised of the \textit{signal misalignment error (${\bm \varepsilon}^{(t)}$)} and \textit{channel nosie}.

Based on \eqref{relation}, we have the following lemma on the signal misalignment error.

\begin{lemma}\label{lemma:aggre_eror_bias_and_mse}
	The statistics of quantized gradient through the over-the-air aggregation are respectively bounded by 
	\begin{align} \quad \label{bias}
		\Vert {\mathbb E}[{\bm \varepsilon}^{(t)}] \Vert^2 &= {\left(\sum_{k=1}^K{\left( 1 - h_{k}^{(t)} p_{k}^{(t)} \right)} \right)}^2 \|{\mathbb E}[{\sf sign}\left({{\bf u}_k^{(t)}}\right)]\|^2 \notag \\ &\leq   {\left(\sum_{k=1}^K{ h_{k}^{(t)} p_{k}^{(t)}- K} \right)}^2 q,
	\end{align}
	and
	\begin{align} \quad \label{MSE}
		{\mathbb E} \left[ \| \bm \varepsilon^{(t)} \|^2 \right] &= \|{\mathbb E}[\bm \varepsilon^{(t)}]\|^2 + {\mathbb E} \left[ \| \bm \varepsilon^{(t)} - {\mathbb E}[\bm \varepsilon^{(t)}]\|^2 \right] \notag \\
		&= {\left(\sum_{k=1}^K{ h_{k}^{(t)} p_{k}^{(t)}-K} \right)}^2 \|{\mathbb E}[{\sf sign}\left({{\bf u}_k^{(t)}}\right)]\|^2 \notag 
		\\ & \qquad \qquad\qquad\qquad\qquad\qquad+ \sum_{i=1}^q\sum_{k=1}^K{\left( h_{k}^{(t)} p_{k}^{(t)}-1 \right)^2} Var[{\sf sign}\left({{\bf u}_k^{(t)}}[i]\right)] \notag 
		\\ &= {\left(\sum_{k=1}^K{h_{k}^{(t)} p_{k}^{(t)}-K} \right)}^2 q + \sum_{k=1}^K{\left( h_{k}^{(t)} p_{k}^{(t)}-1\right)^2} \| {\boldsymbol{\sigma}_{1}} \|^2,
	\end{align}
where we have $\sum_{i=1}^q Var[{\sf sign}\left({{\bf u}_k^{(t)}}[i]\right)] =  \| {\boldsymbol{\sigma}_{1}} \|^2$, ${\boldsymbol{\sigma}_{1}} \in \mathbb{R}^q$. The value of $\| {\boldsymbol{\sigma}_{1}} \|^2$ is bounded by $4q$ due to the bounded values on ${\sf sign}\left({{\bf u}_k^{(t)}}[i]\right)$ and ${\mathbb E}[{\sf sign}\left({{\bf u}_k^{(t)}}\right)]$.
\end{lemma}

As per \eqref{bias} and \eqref{MSE}, the signal misalignment error can be minimized by adjusting the transmission power. We will detail the design of the power control policy in section IV.

\subsection{Convergence Analysis for FEEL with EFOBDA}\label{sec: General_analysis}
 In this section, we provide convergence analysis for the proposed EFOBDA by utilizing the bounded error in Lemma \ref{lemma:error_bound}. We first present the convergence analysis of EFOBDA under general fading channel and then consider simplified AWGN channel. For both scenarios, we compare the results with the scheme without quantization in [25]. 
 \begin{theorem}\label{theo:General}
 	Consider a FEEL system deploying EFOBDA over fading channel, under assumptions 1-4 and learning rate $\eta$, the convergence rate is given by
 	\begin{align} \label{convergence_rate_General}
 		{\mathbb E}\left[\frac{1}{T}\sum_{t=0}^T ||{\bf g}^{(t)}||^2 \right] 
 		\leq \frac{\beta}{\eta} {\left(\frac{F_0-F^*}{ T} + \frac{\eta^2 L B G^2}{\beta^2 } + \frac{\eta^2 L}{2K^2} \sigma_z^2 + \frac{C}{K^2T} \sum_{t=0}^{T}\! \underbrace{{\left(\sum_{k=1}^K{h_{k}^{(t)} p_{k}^{(t)}-K} \right)}^2 q}_{ \Vert {\mathbb E}[{\bm \varepsilon}^{(t)}] \Vert^2} \right.} \notag 
 		\\ {\left. + \frac{\eta^2 L}{2K^2T} \sum_{t=0}^{T}\! {\underbrace{\left( \left(\sum_{k=1}^K{h_{k}^{(t)} p_{k}^{(t)}- K} \right)^2 q + \sum_{k=1}^K {\left( h_{k}^{(t)} p_{k}^{(t)}-1\right)^2} \| {\boldsymbol{\sigma}_{1}} \|^2\right)}_{{\mathbb E} \left[ \| \bm \varepsilon^{(t)} \|^2 \right]}} \right)},
 	\end{align}
 	where $\beta > 0 $ and the scaling factor $B$ and $C$ are given by
 	\begin{align}
 		B = \frac{L (1+\eta) (1-\delta)+ \delta/2}{\rho \delta}, \qquad
 		C = \frac{\eta^2 + \rho^2 \eta +\rho^2}{2\rho}, \notag
 	\end{align}
 	where $\rho > 0$ is the constant used for mean-value inequality.  $\delta$ controls the quantization error as defined in assumption 1. $\Vert {\mathbb E}[{\bm \varepsilon}^{(t)}] \Vert^2$ and ${\mathbb E} \left[ \| \bm \varepsilon^{(t)} \|^2 \right]$ are given in Lemma \ref{lemma:aggre_eror_bias_and_mse}.
 	
 \end{theorem}

\proof See Appendix \ref{app:theo:general}.
\endproof

As observed in Theorem \ref{theo:General}, the upper bound decreases as the number of communication rounds increases until approaching the performance bottleneck regulated by the errors mentioned in \eqref{relation}. The accumulated error on signal misalignment and the variance of channel noise quantify the impact of wireless transmission on the learning performance. The second term in the upper bound, i.e., $\frac{\eta L B G^2}{\beta}$, consists of the weighted quantization error bound, i.e., $\frac{\beta \eta^2 L}{2}\!\!\times\!\!\frac{2(1+\eta) (1-\delta)G^2}{\eta \delta \beta^2}$, and the weighted second-order moment of stochastic gradient, i.e., $\frac{ \eta L}{2\rho\beta} \!\!\times\!\! G^2$. Increasing the error feedback strength $\beta$ reduces the quantization error and the weighted second order moment of stochastic gradient but compromises to the convergence rate. To achieve the optimal convergence performance under the limited communication round $T$, we need to optimize the error-feedback strength $\beta$ to address the trade-off between quantization error and the convergence rate. On the other hand, optimizing the transmission power control policy of edge devices can reduce the aggregation error but under the constraint of modulated error-corrected signal.

As a direct comparison with the scheme without gradient quantization, we introduce the following proposition by rearranging (43) in [\emph{Appendix A}, 25]. 
\begin{proposition}\label{proposition:OPC}
		Consider a FEEL system without gradient quantization over fading channel, under assumptions 1-4 and learning rate $\eta$, the convergence rate is given by
	\begin{align}
		{\mathbb E}\left[\frac{1}{T}\sum_{t=0}^T ||{\bf g}^{(t)}||^2 \right] 
		\leq \frac{1}{\eta(1-\eta)}{\left(\frac{F_0-F^*}{ T} +\frac{\eta^2 L {G}^2}{2}  + \frac{(1+\eta^2 L^2){G}^2}{2TK^2}\sum_{t=0}^T{\left(\sum_{k=1}^K{h_{k}^{(t)} p_{k}^{(t)}-K} \right)}^2 \right.}\notag \\  { \left. + \frac{\eta^2 L }{2 K^2} \sigma_z^2 +\frac{\eta^2 L}{2K^2T} \sum_{t=0}^{T}\! {{\left( \left(\sum_{k=1}^K{h_{k}^{(t)} p_{k}^{(t)}-K} \right)^2 {G}^2 + \sum_{k=1}^K {\left( h_{k}^{(t)} p_{k}^{(t)}-1\right)^2} \| {\boldsymbol{\sigma}} \|^2\right)}} \right)}, 
	\end{align}
\end{proposition}

{\bf Remark 1} (\emph{EFOBDA v.s. Analog modulation}) For comparison, we assume that $\eta \rightarrow 0$, and omit the terms with order higher than $\mathcal{O}(\eta^2)$. The upper bound in Theorem \ref{theo:General} and Proposition \ref{proposition:OPC} are reduced to 1). ${\beta}/{\eta} (\frac{F_0-F^*}{ T} + \frac{\rho q}{2K^2T} (1+{1}/{\eta}) \sum_{t=0}^{T}\! {{(\sum_{k=1}^K{h_{k}^{(t)} p_{k}^{(t)}-K} )}^2})$ and 2). ${1}/{\eta}(\frac{F_0-F^*}{ T} + \frac{{G}^2}{2TK^2}\sum_{t=0}^T{(\sum_{k=1}^K{h_{k}^{(t)} p_{k}^{(t)}-K} )}^2 )$, respectively. Comparing upper bound 1) and 2) leads to the range of $\beta$, i.e., $0< \beta < \min{\{1,\frac{\eta G^2}{\rho q (\eta+1)}\}}$, with which the proposed scheme can achieve a faster convergence than the scheme without gradient quantization. In another word, EFOBDA achieves better performance than OBDA, since the latter is worse than the scheme without quantization as shown in [29].

In the next, we consider AWGN channel and power control $p_k^{(t)}$ is set as $1$ to align the signals transmitted by different devices. The convergence analysis of EFOBDA under AWGN channel is simplified as below.
\begin{corollary}\label{theo:AWGN}
Consider a FEEL system deploying EFOBDA over AWGN channel, under assumptions 1-4 and learning rate $\eta$, the convergence rate is given by
\begin{align} \label{convergence_rate_AWGN}
{\mathbb E}\left[\frac{1}{T}\sum_{t=1}^T ||{\bf g}^{(t)}||^2 \right]\leq \frac{\beta}{\eta(1-\rho/{2 \beta})}\left(\frac{F_0-F^*}{T}+\frac{\eta^2 L D G^2}{ \beta^2}  +\frac{\eta^2 L}{ 2 K^2}{\bm \sigma}_z^2 \right),
\end{align}
where $\rho < 2\beta$ and 
\begin{align}
	D = \frac{ 2L (1+\eta) (1-\delta)+\rho\delta}{ 2 \rho \delta}. \notag
\end{align}
\end{corollary}
\proof See Appendix \ref{app:theo:AWGN}.
\endproof

As discussed in [25], we can avoid the error incurred convergence bottleneck by setting learning rate $\eta = \frac{1}{\sqrt{LT}}$ resulting as
\begin{align} \label{convergence_rate_AWGN_1}
	{\mathbb E}\left[\frac{1}{T}\sum_{t=1}^T ||{\bf g}^{(t)}||^2 \right]\leq \frac{\beta \sqrt{L}}{\sqrt{T}(1-\rho/{2 \beta})}\left({F_0-F^*}+\frac{D G^2}{ \beta^2 \sqrt{T}}  +\frac{{\bm \sigma}_z^2}{ 2\sqrt{T} K^2} \right),
\end{align}

Similarly, by setting $h_k^{(t)} =1 $ and $p_k^{(t)} =1 $ in Proposition \ref{proposition:OPC}, we have the convergence rate of the FEEL system without gradient quantization over AWGN channel.
\begin{corollary}\label{proposition:AWGN_analog_modulation}
	Consider a FEEL system without gradient quantization over AWGN channel, under assumptions 1-4 and learning rate $\eta$, the convergence rate is given by
	\begin{align}
		{\mathbb E}\left[\frac{1}{T}\sum_{t=1}^T ||{\bf g}^{(t)}||^2 \right] \leq \frac{1}{\eta}\left(\frac{F_0-F^*}{ T} +\frac{\eta^2 L {G}^2}{2} + \frac{\eta^2 L }{2 K^2} {\bm \sigma}_z^2  \right), 
	\end{align}
\end{corollary}
\vspace{4mm}

{\bf Remark 2} Compare Corollary \ref{theo:AWGN} and Corollary \ref{proposition:AWGN_analog_modulation}, when $\eta \rightarrow 0$ and $ 0 <\frac{1-\sqrt{1-2\rho}}{2}< \beta < \frac{1+\sqrt{1-2\rho}}{2} < 1$, the proposed EFOBDA converges faster than the scheme without gradient quantization.

\section{power control optimization}

In this section, we will present the power control optimization policy to minimize the upper bound in Theorem \ref{theo:General}.

\subsection{Problem Formulation}

To start with, we first formulate the optimization problem by minimizing the upper bound in \eqref{convergence_rate_General} under power constraint \eqref{power_con}. Since that ${F_0-F^*}/{ T}$, ${\eta^2 L B G^2}/{\beta^2 }$ and ${\eta^2 L \sigma_z^2} / {2K^2} $ in \eqref{convergence_rate_General} are irrelevant to the power control policy, minimizing the upper bound over $ {\{p_{k}^{(t)} \}} $ is equivalent as minimizing the misalignment error $ \Phi({\{p_{k}^{(t)} \}}) $.
\begin{align}\label{objective}
	\!\!\! \Phi({\{p_{k}^{(t)} \}}) = \frac{(\rho^2 \!+\! \rho^2 \eta \!+\! \eta^2 (\rho L +1))q }{2\rho TK^2} \sum_{t=0}^{T-1}  {\left(\!\sum_{k=1}^K\!\!{h_{k}^{(t)} p_{k}^{(t)}-K\!\!} \right)}^2 \! \qquad \qquad \qquad \qquad \notag \\ + \! \frac{\eta^2 L \| {\boldsymbol{\sigma}_{1}} \|^2 }{2TK^2} \! \sum_{t=0}^{T-1} \! \sum_{k=1}^K \!\!{\left( h_{k}^{(t)} p_{k}^{(t)}-1\right)^2}. 
\end{align}

Then the optimal problem is formulated as 
\begin{align}\label{P1}
	\boldsymbol {\mathrm P1}:&\underset{\{p_k^{(t)} \geq 0 \}}{\min} \,\, \Phi({\{p_k^{(t)} \}}) \notag \\ &s.t.\quad {|p_k^{(t)}|^2} \leq \frac{P_0}{M}, \quad \forall k \in {\mathcal{K}}
\end{align}

It is observed that $\boldsymbol {\mathrm P1}$ can be solved as $T$ parallel optimizations and it is equivalent to focus on the $t$-th iteration as
\begin{align}\label{P2}
	\boldsymbol {\mathrm P2}:&\underset{\{p_k^{(t)}\}_{k=1}^K}{\min} \,\, \frac{(\rho^2 \!+\! \rho^2 \eta \!+\! \eta^2 (\rho L +1))q }{2\rho TK^2} {\left(\!\sum_{k=1}^K\!\!{h_{k}^{(t)} p_{k}^{(t)}-K\!\!} \right)}^2 \! + \! \frac{\eta^2 L \| {\boldsymbol{\sigma}_{1}} \|^2 }{2TK^2} \sum_{k=1}^K \!\!{\left( h_{k}^{(t)} p_{k}^{(t)}-1\right)^2}, t=0,...,T \notag \\ &s.t.\quad {|p_k^{(t)}|^2} \leq \frac{P_0}{M}, \quad \forall k \in {\mathcal{K}}
\end{align}

\subsection{Optimal Solution}
The problem $ \boldsymbol {\mathrm P2} $ is seen to be convex and the close-form solution involves applying the Lagrange method and Karush-Kuhn-Tucker (KKT) conditions. The optimal solution to problem $ \boldsymbol {\mathrm P1} $ is given by
\begin{align}\label{op_solution}
	{p_{k}^{(t)}}^*= \frac{A h_k^{(t)}}{{h_k^{(t)}}^2 + \frac{2}{\eta^2 L \| {\boldsymbol{\sigma}_{1}} \|^2} {\lambda_k^{(t)}}^*},
\end{align}
where 
\begin{align}
	A = \frac{ \rho \eta^2 L \| {\boldsymbol{\sigma}_{1}} \|^2 + \left(\left(\rho L+1\right)\eta^2+\rho^2 \eta +\rho^2\right)q K}{ \eta^2 L \| {\boldsymbol{\sigma}_{1}} \|^2 \left( \rho + {\left(\left(\rho L+1\right)\eta^2+\rho^2 \eta +\rho^2\right)q \sum_{j=1}^K {\frac{h_j^{(t)}}{B_j} }} \right)}
\end{align}
and
\begin{align}
	B_j ={\eta^2 L \| {\boldsymbol{\sigma}_{1}} \|^2 h_{j}^{(t)}} + \frac{2{\lambda_j^{(t)}}^*}{h_{j}^{(t)}}.
\end{align}
and ${\lambda_k^{(t)}}^*$ should satisfy that
\begin{align}
	\sum_{k=1}^{K} {{\lambda_k^{(t)}}^* \left(|p_k^{(t)}|^2-\frac{P_0}{M}\right)} = 0.
\end{align}

\proof See Appendix \ref{solution}.
\endproof

{\bf Remark 3} (\emph{The effect of power constraint.})
As observed from \eqref{op_solution}, the optimal solution shows a \emph{regularized channel inversion} structure with a regularization term $ \frac{2\lambda_k^* }{\eta^2 L \| {\boldsymbol{\sigma}_{1}} \|^2} $ related to optimal dual variable $ \lambda_k^* $. The scaling factor $ A $ is determined by channel coefficients and optimal dual variables associated with all edge devices such that it is the same for different edge devices. Especially, when the power constraint is less stringent, all the dual variables become zero, the optimal power scaling strategy reduces to the channel inversion policy.

\section{Simulation results}\label{simulation}

In this section, we evaluate the accuracy and convergence performance of the proposed scheme. We consider a FEEL system with one edge server and $K = 20$ edge devices. The sub-channel coefficients over different communication rounds are i.i.d. Rayleigh distributed, i.e., $ h_k[n,m] \sim \mathcal{CN}\left( 0, 1\right)$. The average receive SNR is set to be 10 dB unless specified otherwise. The learning task of numerical experiments is image classification using the well-known MNIST and CIFAR10 datasets respectively. The MNIST datasets consist of $10$ classes of black-and-white digits ranging from ``$0$" to ``$9$". The corresponding classifier model is implemented using a $6$-layer \emph{convolution neural network} (CNN)  that consists of two $5\times5$ convolution layers with ReLU activation, each followed with a $2\times2$ max pooling; a fully connected layer with $512$ units,  ReLU activation; and a final soft-max output layer. CIFAR10 consists of $10$ classes of $ 32 \times 32 $ RGB color images. For the CIFAR10 datasets, the well-known classifier model, ResNet18 with batch normalization proposed in \cite{he2016deep}, is applied. {In the experiments, we consider non-i.i.d MNIST datasets and i.i.d CIFAR10 datasets. The learning rate $\eta$ is set within the range $(0.001,0.1)$.}

For performance comparison, we consider the following four benchmark schemes:
\begin{itemize}
	\item {{\bf BAA }\cite{zhu2019broadband}}: Edge device transmits the local gradient by analog modulation without quantization.
	
	\item {{\bf BAA with optimized power control (BAA+OPC)}}: Edge device transmits the local gradient by analog modulation without quantization with an optimized power control policy.

	\item {{\bf OBDA}\cite{zhu2020one}}: Each device transmits the one-bit quantized gradient with \emph{truncated channel inversion} power control policy.
	
	\item {\bf OBDA with optimized power control (OBDA+OPC)}: Each device transmits the one-bit quantized gradient with an optimized power control policy in terms of minimizing the aggregation error.
\end{itemize}

\subsection{Performance Evaluation of EFOBDA}
\begin{figure}[t]
	\centering
	\subfigure[Test accuracy versus $ T $ over AWGN MAC]{
		\includegraphics[width=8cm]{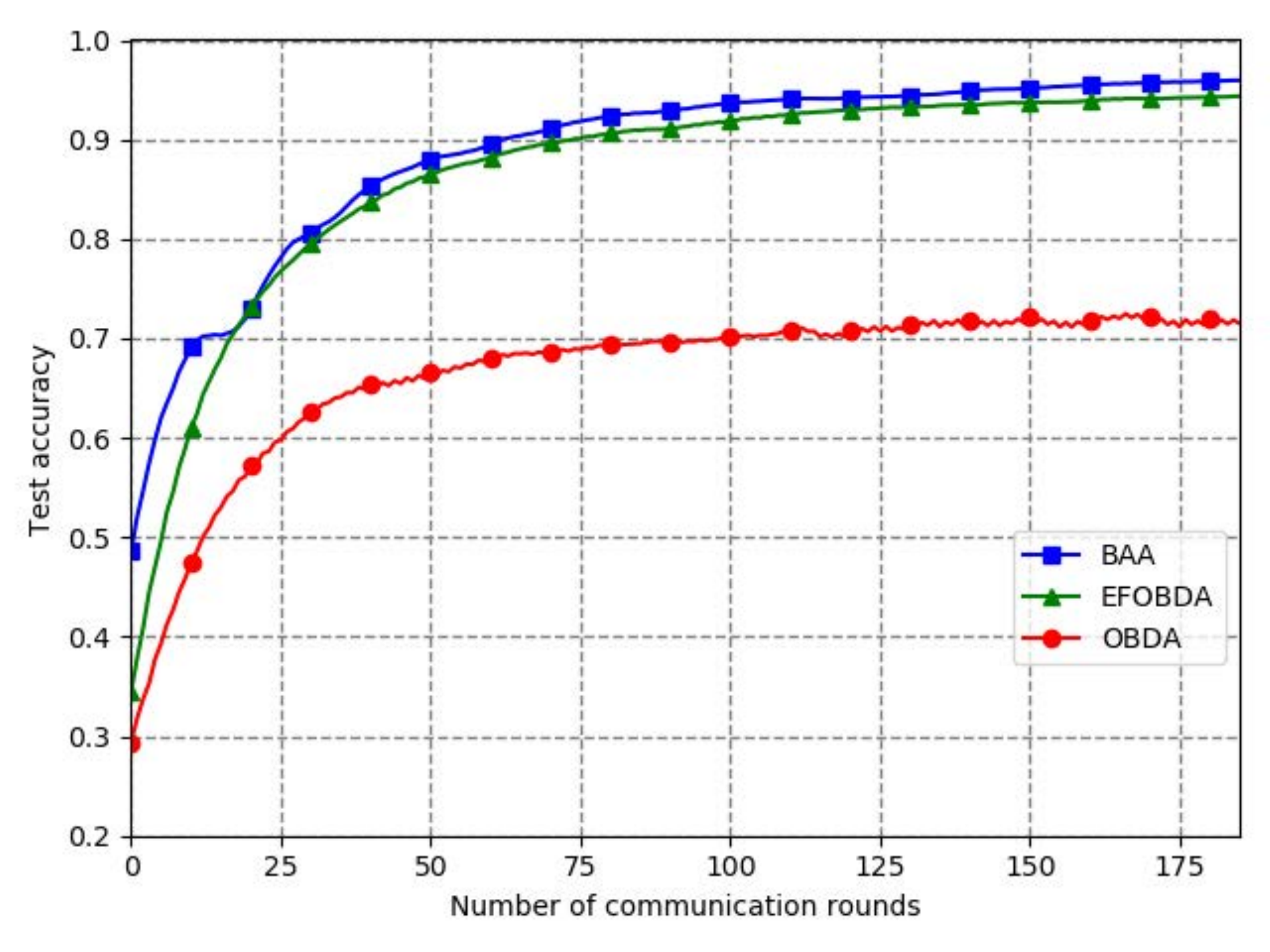}
	}
	\subfigure[Test accuracy versus $ T $ over fading MAC]{		
		\includegraphics[width=8cm]{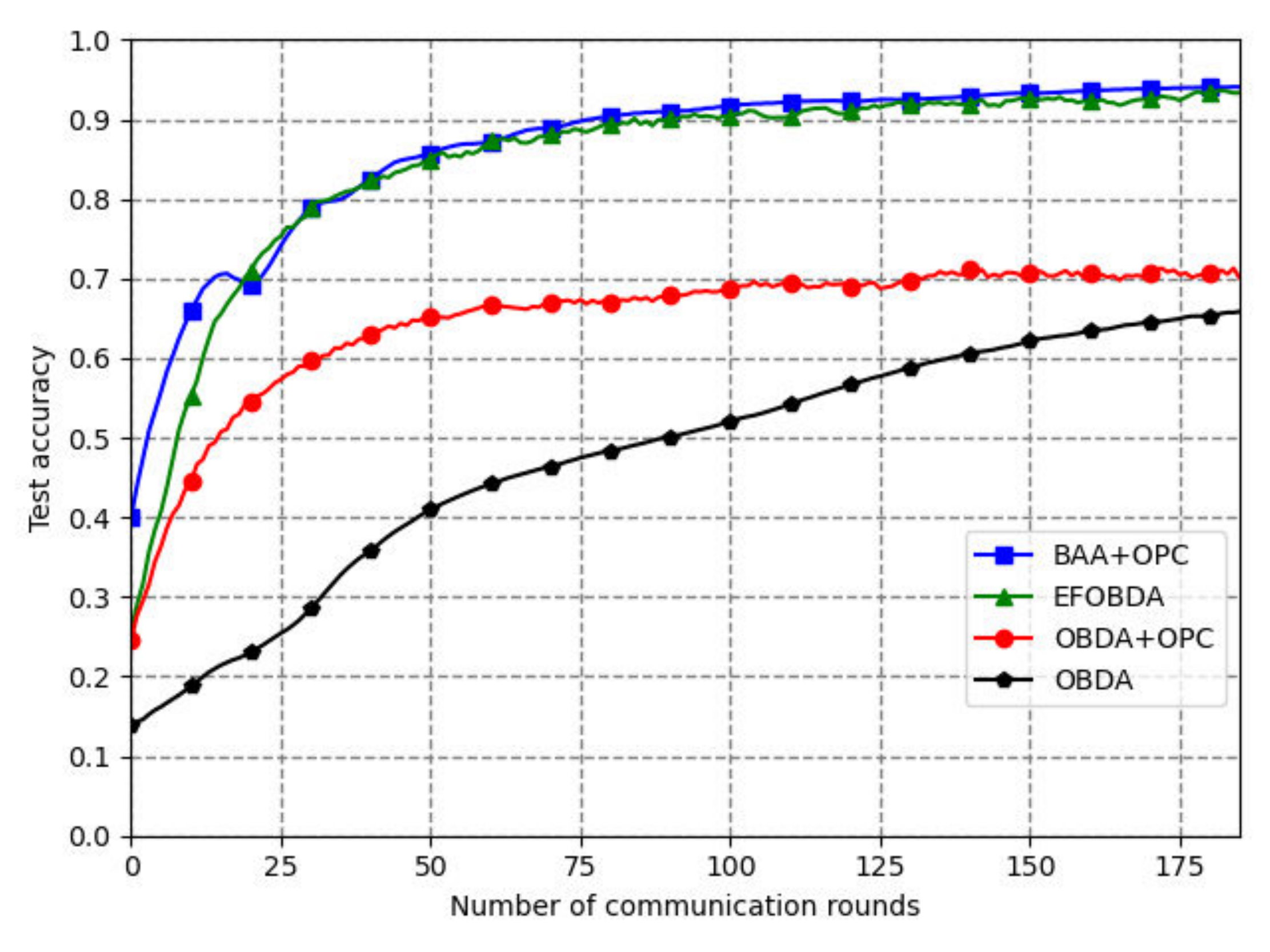}
	}
	\caption{Convergence performance in two scenarios}
	\label{mnist_per_val}
	\vspace{-4mm}
\end{figure}

\begin{figure}[t]
	\centering
	\subfigure[Test accuracy versus $ T $]{
		\includegraphics[width=8cm]{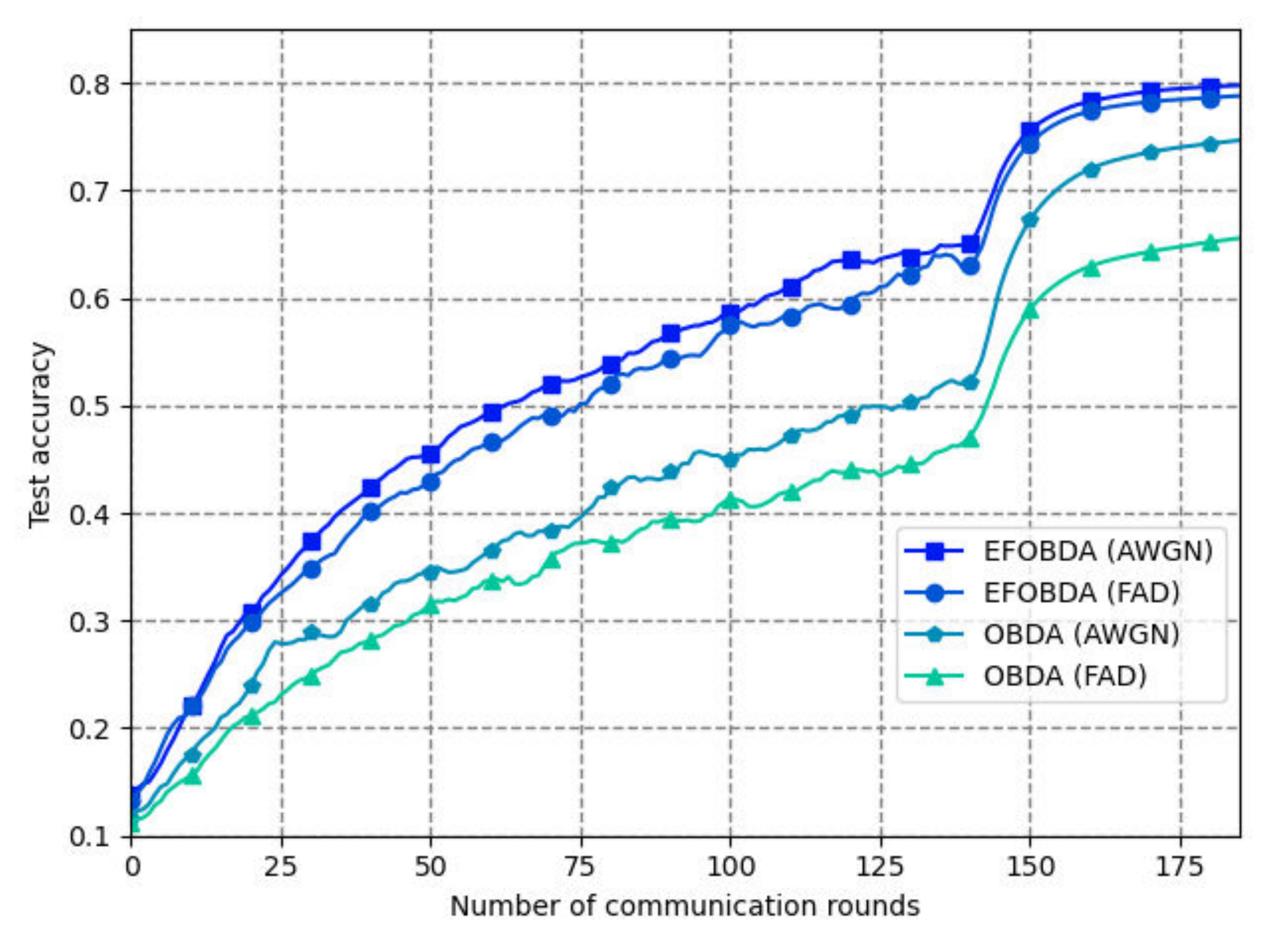}
	}
	\subfigure[Train loss versus $ T $]{		
		\includegraphics[width=8cm]{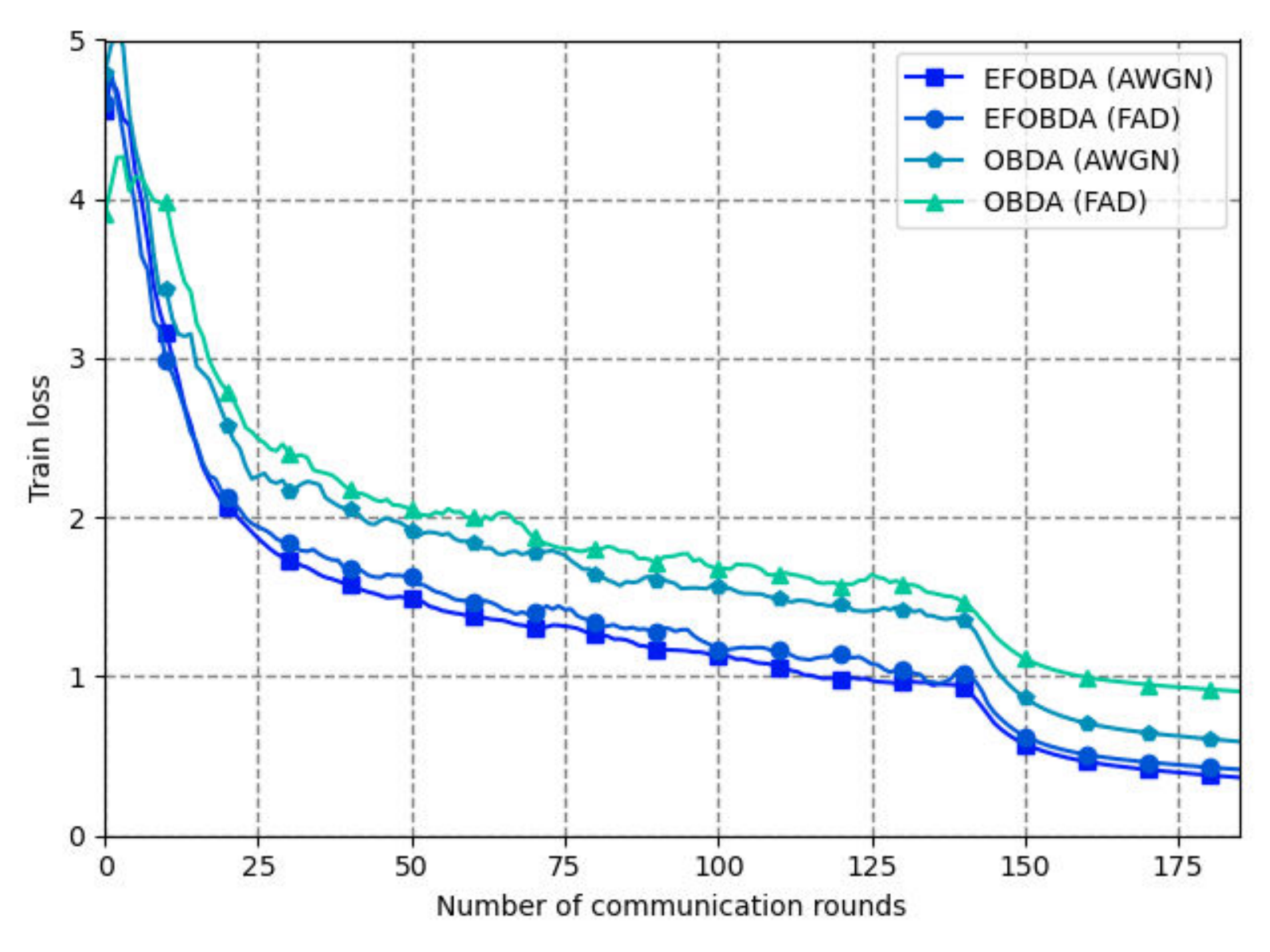}
	}
	\caption{Convergence performance on Cifar datasets}
	\label{cifar}
	
	\vspace{-4mm}
\end{figure}

For both MNIST and CIFAR10 datasets, the effectiveness of EFOBDA is evaluated in the two considered scenarios, namely over an AWGN MAC, and fading MAC with perfect CSI. Test accuracy and train loss are plotted as functions of the number of communication rounds in Fig. \ref{mnist_per_val} and Fig. \ref{cifar}. First, the proposed scheme is observed to achieve nearly the same convergence rate and performance as BAA and performs better than OBDA in Fig. \ref{mnist_per_val}(a). This is because the loss gradient information induced by one-bit quantization can be transmitted by error-feedback, which is almost equivalent to the case without quantization. Secondly, it is observed from Fig. \ref{mnist_per_val}(b) that EFOBDA significantly outperforms OBDA with \emph{truncated channel inversion} power control policy over fading MAC. This is because the power control policy is optimized to address the aggregation error incurred by fading MAC while \emph{truncated channel inversion} is an heuristic power control policy without further optimization.

\subsection{Effect of error-feedback strength}
\begin{figure}[t]
	\centering
	\subfigure[Test accuracy versus error-feedback strength (AWGN)]{
		\includegraphics[width=8cm]{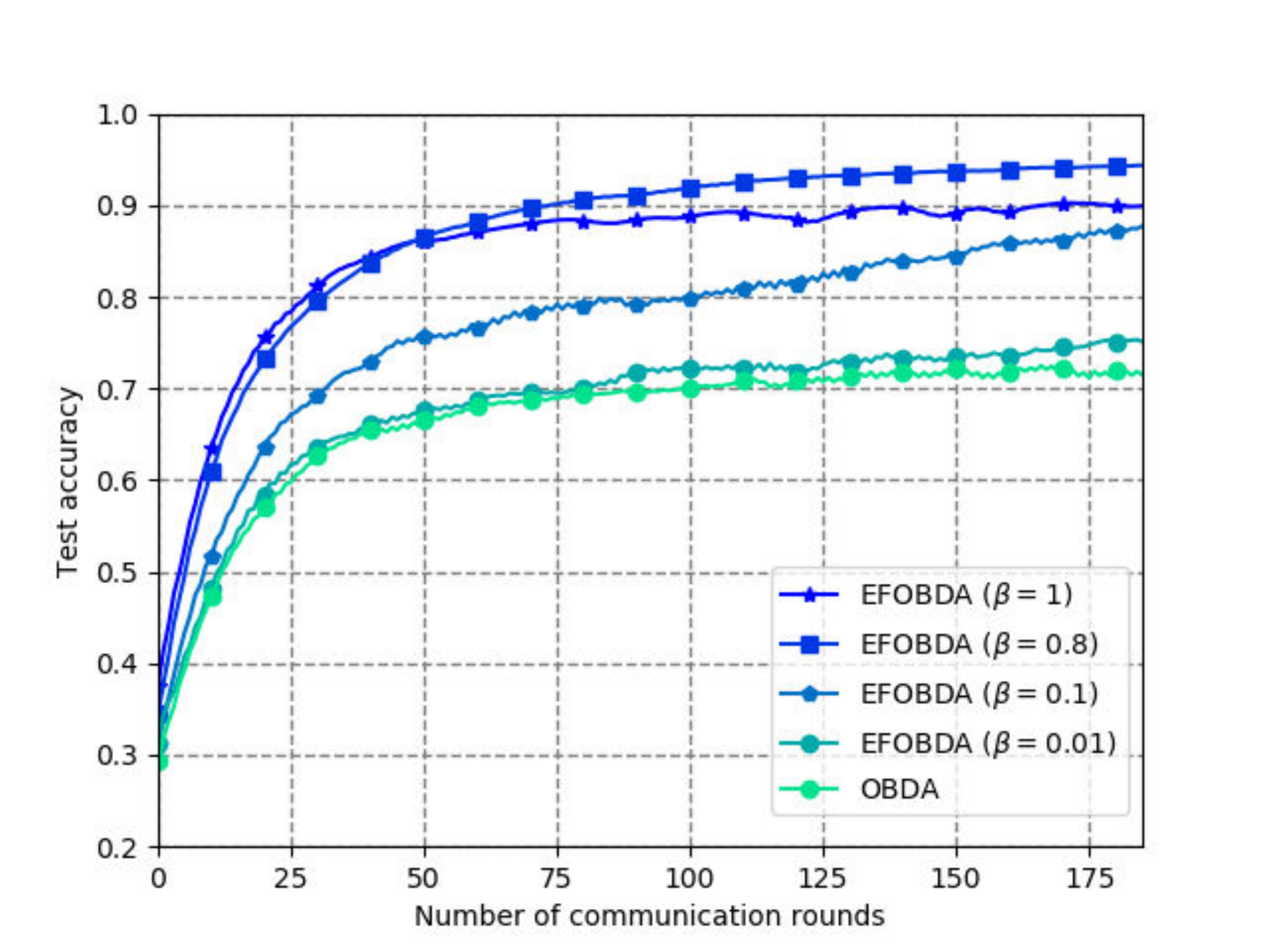}
	}
	\subfigure[Test accuracy versus error-feedback strength (FAD)]{		
		\includegraphics[width=8cm]{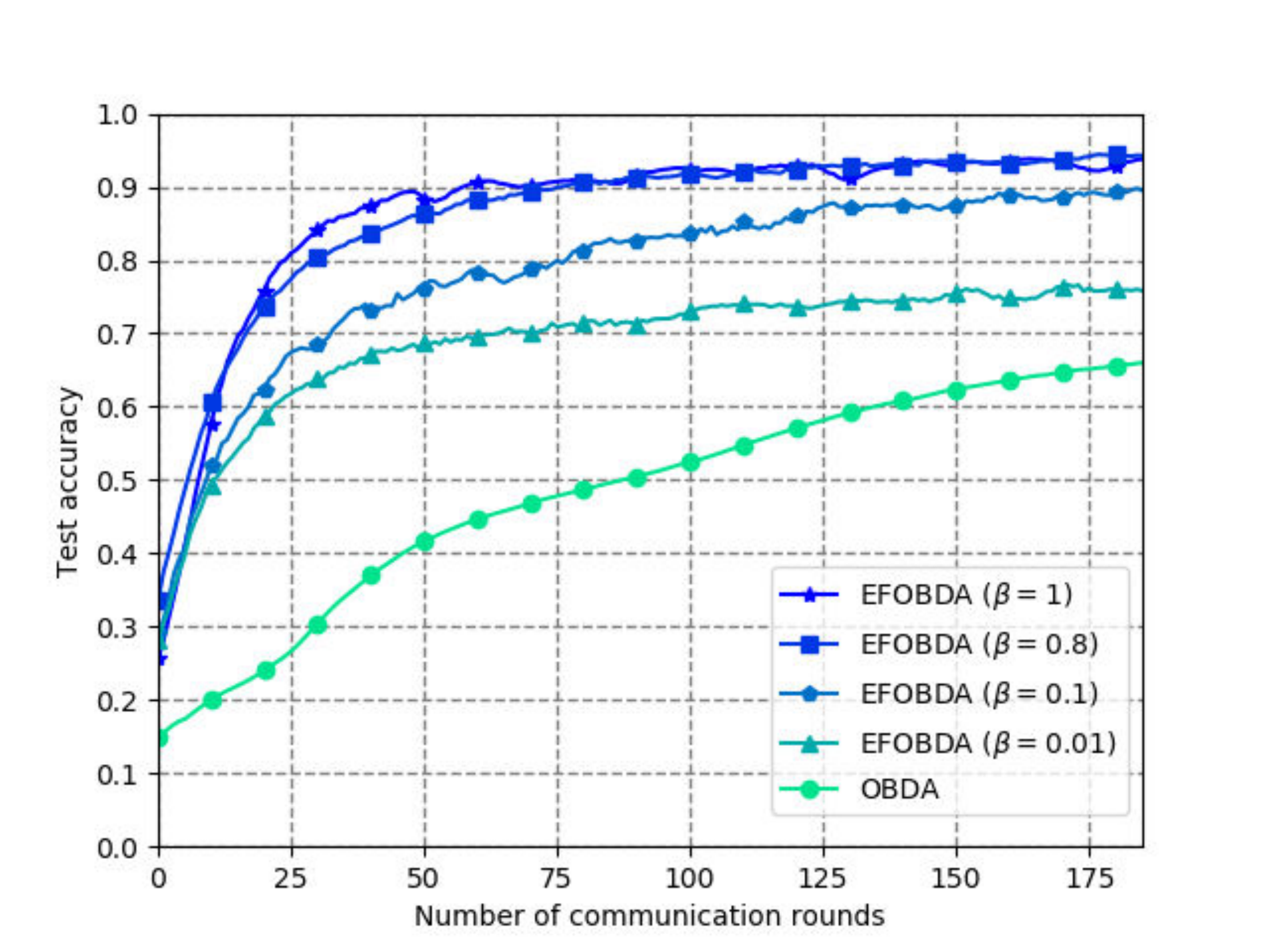}
	}
	\subfigure[Train loss versus error-feedback strength (AWGN)]{
	\includegraphics[width=8cm]{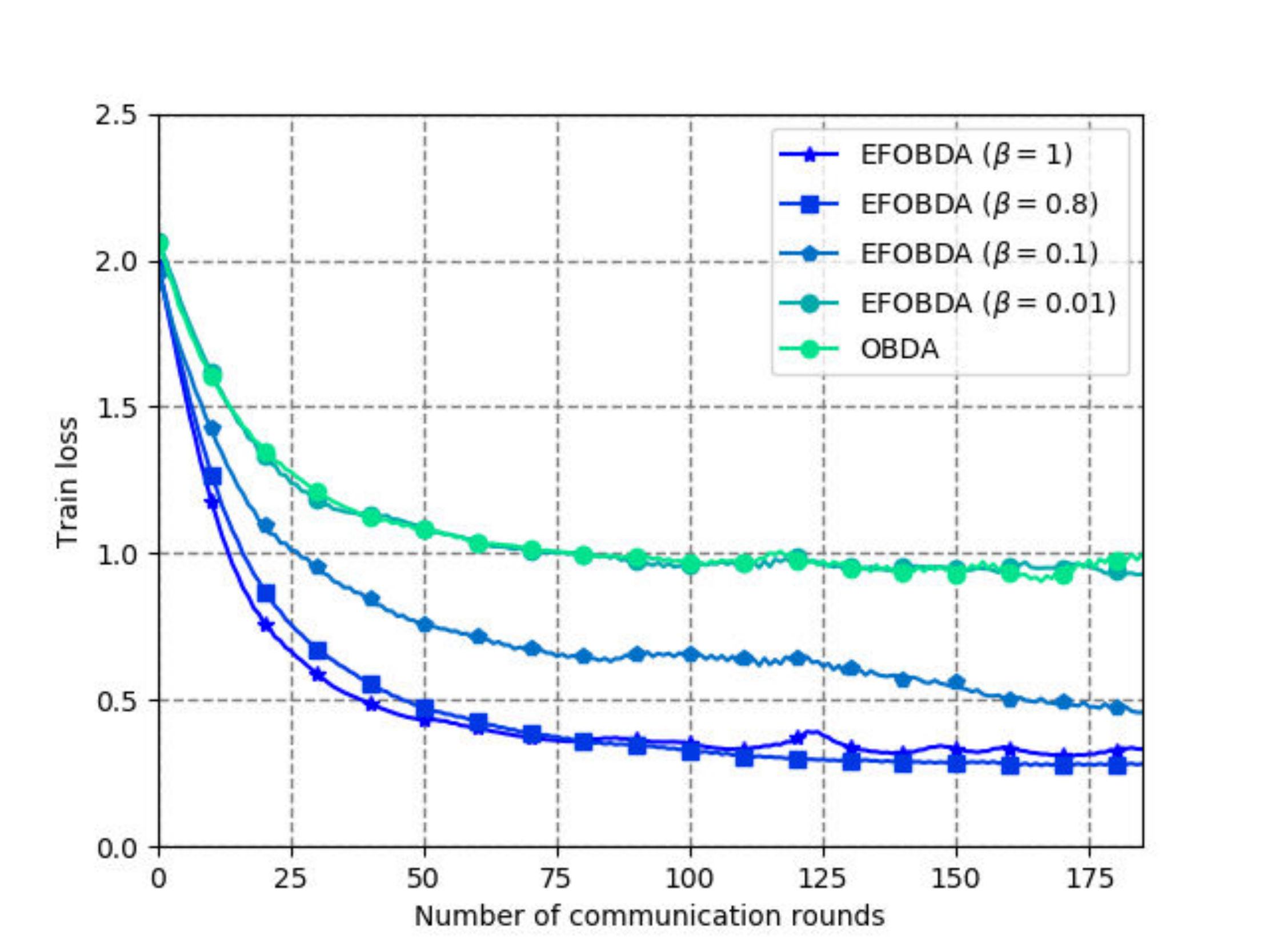}
	}
	\subfigure[Train loss versus error-feedback strength (FAD)]{		
	\includegraphics[width=8cm]{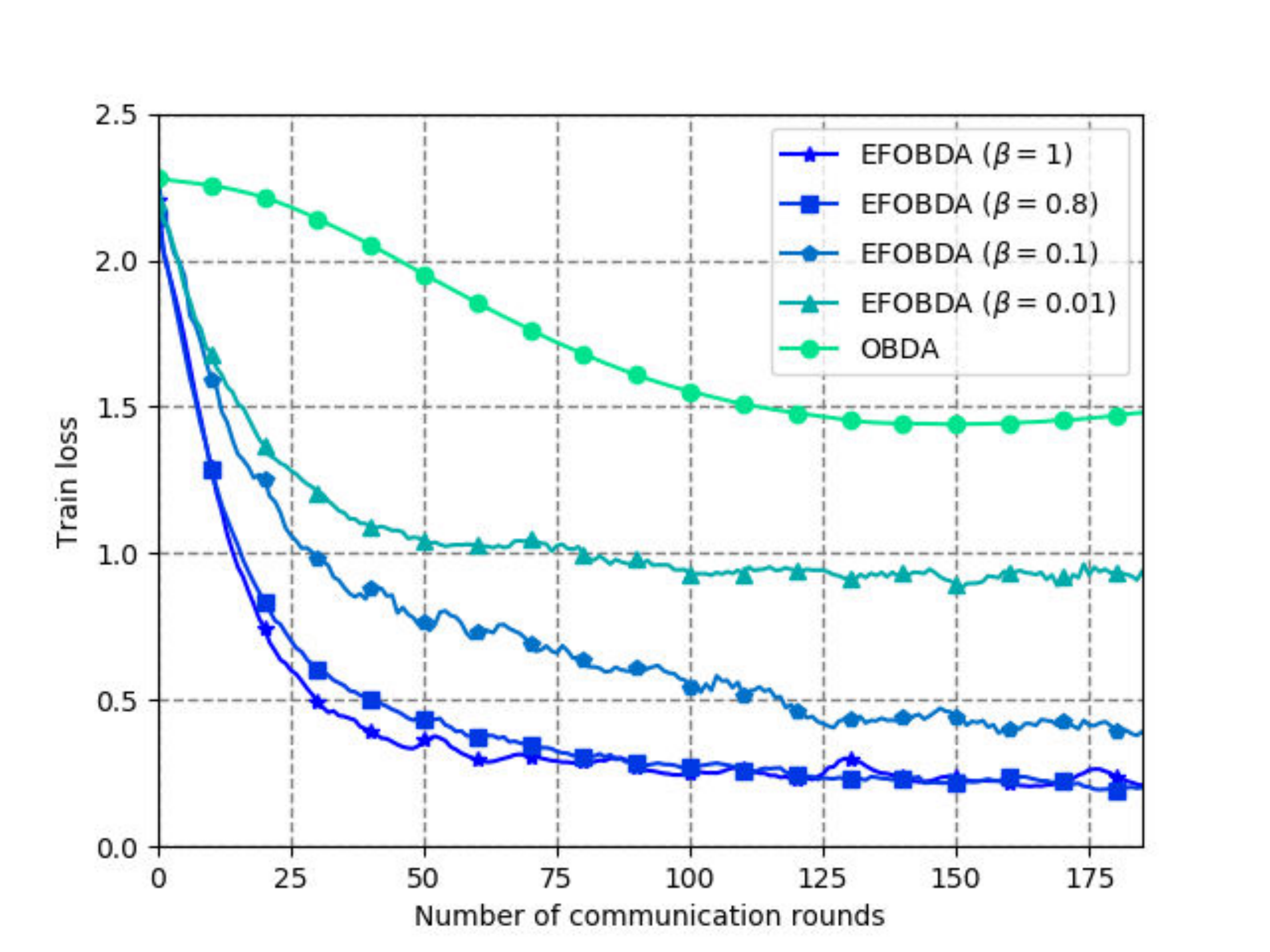}
	}
	\caption{Effect of error-feedback strength}
	\label{mnist_beta}
	\vspace{-4mm}
\end{figure}
To evaluate the impact of error-feedback strength on the performance of EFOBDA, we consider four cases where the value of $ \beta $ takes $ 0.01, 0.1, 0.8, 1 $, respectively. First, it is observed from Fig.~\ref{mnist_beta} that the test accuracy and training loss of EFOBDA vary with $ \beta $. For total communication round $T=180$,  $\beta=0.8$ achieves best performance. Fig. \ref{mnist_beta}(a) also shows that when the value of $ \beta $ is sufficiently small, e.g., $ \beta=0.01 $, the convergence rate and performance of EFOBDA over an AWGN MAC is nearly the same as OBDA. Secondly, in the presence of power control policy over fading MAC, the performance of EFOBDA cannot reduce to OBDA when $ \beta=0.01 $ as shown in Fig. \ref{mnist_beta}(b). This shows the benefit of power control optimization in improving the learning performance. Furthermore, the proposed scheme is observed to converge faster at the beginning of training when $ \beta>0.8 $ but perform worse in the end. This indicates the trade-off between convergence rate and error as discussed in Theorem \ref{theo:General}.

\subsection{Effect of the Number of Devices}
\begin{figure}[t]
	\centering
	{
		\includegraphics[width=8cm]{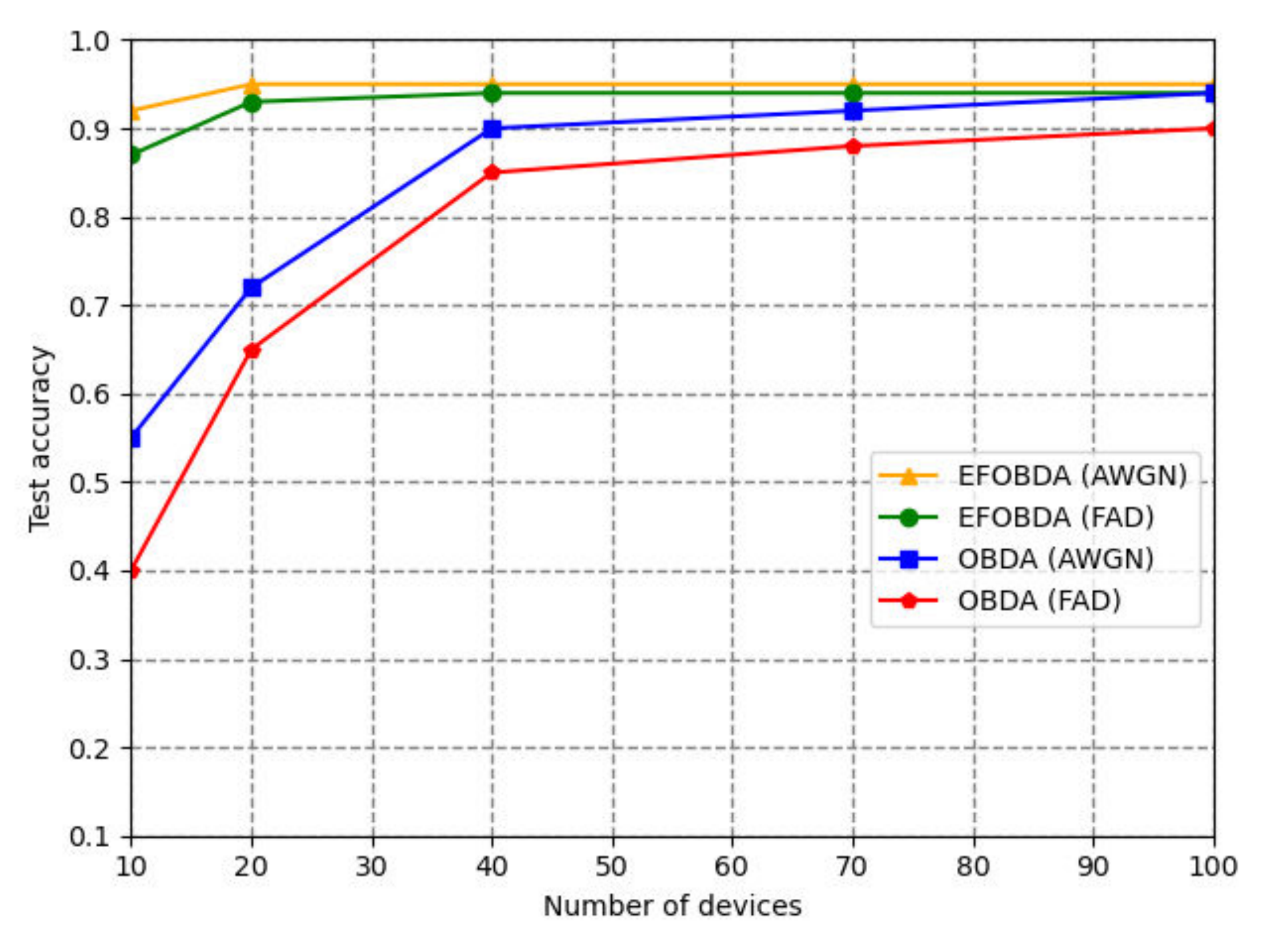}
	}
	\caption{Effect of the Number of Devices}
	\label{mnist_k}
	\vspace{-4mm}
\end{figure}

\begin{figure}[t]
	\centering
	{		
		\includegraphics[width=8cm]{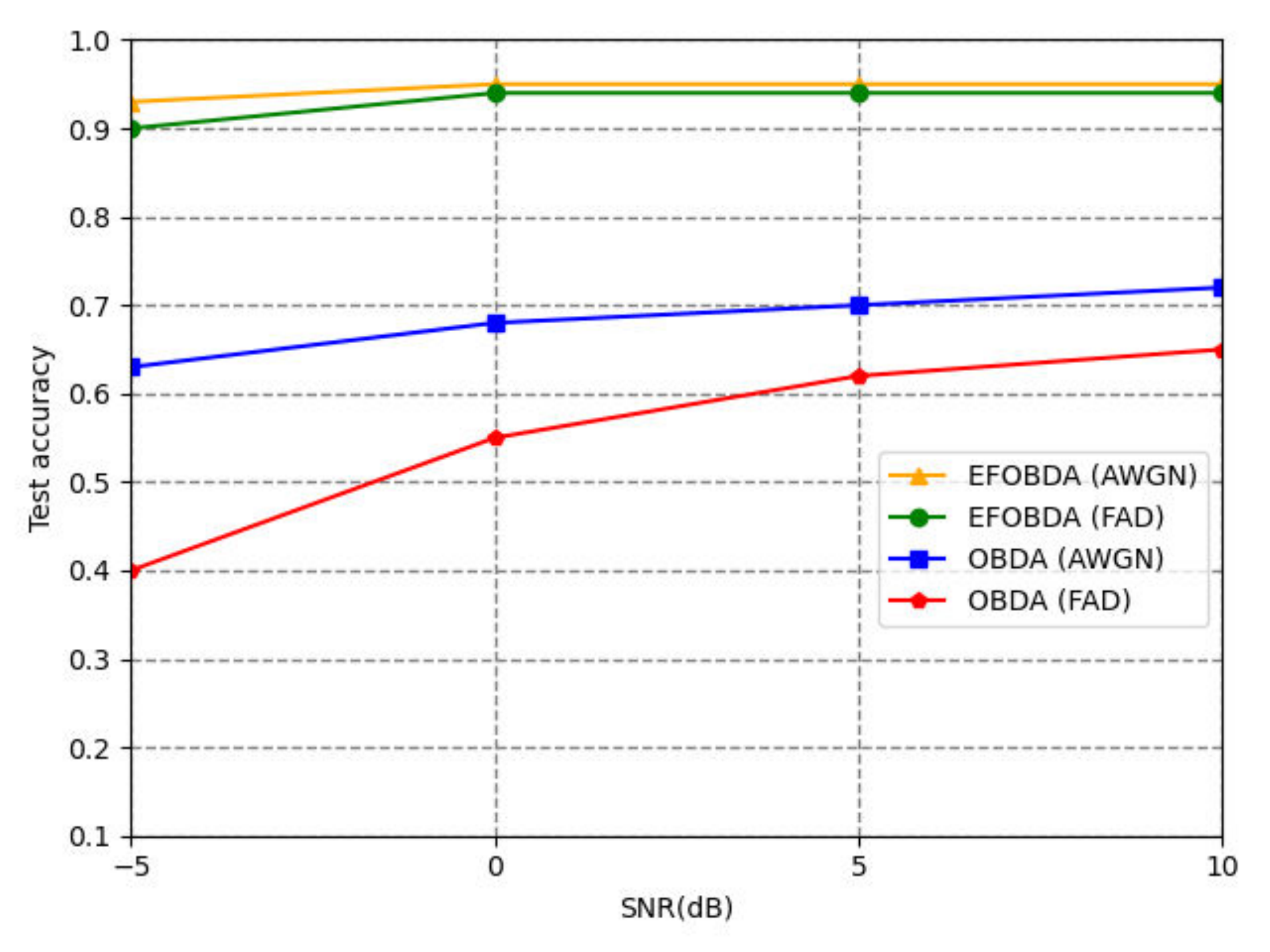}
	}
	\caption{Effect of SNR}
	\label{mnist_snr}
	\vspace{-4mm}
\end{figure}

The effect of the number of the devices on the convergence is illustrated in Fig. \ref{mnist_k}. First, it is observed that the test accuracy increase with $ K $. This is because the impact of signal misalignment error and channel noise on the convergence rate is decreasing with a larger device population as shown in Theorem \ref{theo:General} and Corollary \ref{theo:AWGN}. Secondly, we fortunately find that the test accuracy of EFOBDA decreases at lower rates as the decreases of device population than OBDA. This indicates that the improvement of EFOBDA compared with OBDA is mainly shown in the case with a small number of devices.

\subsection{Effect of SNR}
The effect of SNR on the convergence is illustrated in Fig. \ref{mnist_snr}. First, it is observed that the test accuracy increase with SNR for both EFOBDA and OBDA. This is because the impact of channel noise on the convergence rate is decreasing with a larger SNR as shown in Theorem \ref{theo:General} and Corollary \ref{theo:AWGN}. Secondly, it is observed that the test accuracy of EFOBDA decreases at lower rates as the decreases of SNR than OBDA. This indicates that the improvement of EFOBDA compared with OBDA is mainly shown in the case with a small SNR.

\section{Conclusion}

In this paper, we have proposed a new digital over-the-air gradient aggregation scheme, called EFOBDA, by introducing error feedback to one-bit SGD. To evaluate its performance, we first analyzed the convergence behavior of EFOBDA and characterized the effect of error-feedback, w.r.t. error-feedback strength, and the impact of aggregation errors over the whole training period. The convergence results show that under specific settings of the error-feedback parameters, the proposed scheme achieves better performance than the scheme without quantization in [25], from where, it also justifies that error-feedback can improves the performance of OBDA in [29]. Next, we intended to optimize the convergence rate over the power control parameters. The optimal solutions are observed to follow the regularized channel inversion structures. Finally, the numerical results also indicated that the proposed scheme achieved a significantly faster convergence rate and better performance compared with benchmark scheme without error-feedback and optimal power control policy. For the future work, there are still a lot of interesting issues that are worth investigating. We will consider the generalization of the current work to more complicated channel scenarios, e.g., fading MAC without perfect CSI, where new power control policies should be designed to alleviate the channel estimation error. The investigation of the proposed scheme with the non-i.i.d. data is also an interesting future research direction.

\clearpage
\appendix
\subsection{Proof of Theorem 1}\label{app:theo:general}
The proof follows the widely-adopted strategy that relates the norm of the gradient to the expected improvement of objective at each communication round which composes the total possible improvement under Assumption 3. The key technical challenge we overcome is to analyze the effect of error-feedback with a biased gradient compression. The proof can be extended to more complicated scenarios as detailed in the sequel.

Let $\hat{\bf w}^{(t)}={\bf w}^{(t)} - \eta \frac{1}{K} \sum_{k=1}^K{{\bf e}_k^{(t)}}$ such that $\hat{\bf w}^{(t)}$ is updated in the same way as ${\bf w}^{(t)}$ in the non error-feedback scenario. Then we have the following lemma about $\hat{\bf w}^{(t)}$.
\begin{lemma}\label{lemma:recurence}
	Let $\hat{\bf w}^{(t)}={\bf w}^{(t)} - \eta \frac{1}{K} \sum_{k=1}^K{{\bf e}_k^{(t)}}$, we have
	\begin{align} \label{eq:xt}
		\hat{\bf w}^{(t+1)} = \hat{\bf w}^{(t)}-\frac{\eta}{\beta K} \sum_{k=1}^K{{{\bf g}}_k^{(t)}} - \frac{\eta}{K} {\boldsymbol \varepsilon}^{(t)} - \frac{\eta}{K} {\bf z}.
	\end{align}
\end{lemma}
\proof See Appendix \ref{app:lemma 3}.
\endproof

According to Assumption 3, the expected improvement of objective in a single communication round is upper bounded by:
\vspace{-4mm}
\begin{align}\label{eq:expected_improvement}
	&{\mathbb E}\left[F(\hat{\bf w}^{(t+1)})-F(\hat{\bf w}^{(t)})\right] \notag \\ &\leq \nabla {F(\hat{\bf w}^{(t)})}^T {\mathbb E}\left(\hat{\bf w}^{(t+1)}-\hat{\bf w}^{(t)}\right)+\frac{L}{2} {\mathbb E}||\hat{\bf w}^{(t+1)}-\hat{\bf w}^{(t)}||^2,\notag \\ &\leq \left[\nabla F(\hat{\bf w}^{(t)})-{\bf g}^{(t)}\right]^T {\mathbb E}\left(\hat{\bf w}^{(t+1)}-\hat{\bf w}^{(t)}\right)+\frac{L}{2} {\mathbb E}||\hat{\bf w}^{(t+1)}-\hat{\bf w}^{(t)}||^2 + {{\bf g}^{(t)}}^T {\mathbb E}\left(\hat{\bf w}^{(t+1)}-\hat{\bf w}^{(t)}\right).
\end{align} 
The first term on the right side of \eqref{eq:expected_improvement} is bounded as shown in the following.
\begin{align}
	\left[\nabla F(\hat{\bf w}^{(t)})-{\bf g}^{(t)}\right]^T {\mathbb E}\left(\hat{\bf w}^{(t+1)}-\hat{\bf w}^{(t)}\right) &= \eta \left[\nabla F(\hat{\bf w}^{(t)})-{\bf g}^{(t)}\right]^T \left(-\frac{{\bf g}^{(t)}}{\beta} - \frac{{\mathbb E}\left(\varepsilon^{(t)}\right)}{K} \right),\notag  
	\\ &\leq \frac{\eta \left[\nabla F(\hat{\bf w}^{(t)})-{\bf g}^{(t)}\right]^2}{2 \rho} + \frac{\eta \rho}{2} \left(\frac{{\bf g}^{(t)}}{\beta}+ \frac{{\mathbb E}\left(\varepsilon^{(t)}\right)}{K}\right)^2.
\end{align}
where the inequity follows from the mean-value inequality and holds for any $\rho > 0$. To bound $ \left[\nabla F(\hat{\bf w}^{(t)})-{\bf g}^{(t)}\right]^2 $, we use an alternate definition of smoothness of loss function:
\begin{align}
	\|{\nabla F}({\bf w}') - {\nabla F}({\bf w})\| \leq L \|{\bf w}' - {\bf w}\|.
\end{align}
We continue as:
\begin{align}
	\left[\nabla F(\hat{\bf w}^{(t)})-{\bf g}^{(t)}\right]^T {\mathbb E}\left(\hat{\bf w}^{(t+1)}-\hat{\bf w}^{(t)}\right)&\leq \frac{\eta L^2}{2\rho} ||\hat{\bf w}^{(t)} - {\bf w}^{(t)}||^2  + \frac{\eta \rho}{2} \left(\frac{{\bf g}^{(t)}}{\beta}+ \frac{{\mathbb E}\left(\varepsilon^{(t)}\right)}{K}\right)^2  ,\notag 
	\\ &\leq \frac{\eta L^2}{2\rho} ||\frac{\eta}{K} \sum_{k=1}^K{{\bf e}_k^{(t)}}||^2 + \frac{\eta \rho}{2} \left(\frac{{\bf g}^{(t)}}{\beta}+ \frac{{\mathbb E}\left(\varepsilon^{(t)}\right)}{K}\right)^2.  	
\end{align}
To bound the quantization error, we have the following lemma.

\begin{lemma}\label{lemma:error_bound}
	The quantization error is bounded by:
	\begin{align}
		{\mathbb E}||\frac{1}{K} \sum_{k=1}^K{{\bf e}_k^{(t)}}||^2 \leq \frac{2(1+\eta) (1-\delta)G^2}{\eta \delta \beta^2}, \quad \forall k,t.
	\end{align}
\end{lemma}
\proof See Appendix \ref{app:lemma 4}.
\endproof

Thus the first term on the right side of \eqref{eq:expected_improvement} is bounded by
\begin{align}\label{term1}
	&\left[\nabla F(\hat{\bf w}^{(t)})-{\bf g}^{(t)}\right]^T {\mathbb E}\left(\hat{\bf w}^{(t+1)}-\hat{\bf w}^{(t)}\right) \notag \\ 
	&\qquad\qquad \leq \frac{\eta^2 L^2 (1+\eta) (1-\delta)G^2}{\rho \delta \beta^2}  + \frac{\eta \rho}{2 \beta^2} || {\bf g}^{(t)}||^2 + \frac{\eta \rho}{2K^2}||{\mathbb E}\left(\varepsilon^{(t)}\right)||^2 + \frac{\eta \rho}{\beta K}{{\bf g}^{(t)}}^T {\mathbb E}\left(\varepsilon^{(t)}\right).
\end{align}

Next, we can bound the second term as follows.
\begin{align}\label{term2}
	\frac{L}{2} {\mathbb E}||\hat{\bf w}^{(t+1)}-\hat{\bf w}^{(t)}||^2 \leq \eta^2 \left(\frac{G^2}{\beta^2}+ \frac{{\mathbb E}\left(||\varepsilon^{(t)}||^2\right)}{K^2}+\frac{\sigma_z^2}{K^2} + \frac{2}{\beta K} {{\bf g}^{(t)}}^T {\mathbb E}\left(\varepsilon^{(t)}\right) \right).
\end{align}
Plugging \eqref{term1} and \eqref{term2} into \eqref{eq:expected_improvement} yields
\begin{align}
	{\mathbb E}\left[F(\hat{\bf w}^{(t+1)})-F(\hat{\bf w}^{(t)})\right] 
	\leq \frac{\eta}{\beta} \left(\frac{\rho}{2\beta}-1\right)||{\bf g}^{(t)}||^2  + \frac{\eta}{K} \left(\frac{\eta L +\rho}{\beta}-1\right) {{\bf g}^{(t)}}^T {\mathbb E}\left(\varepsilon^{(t)}\right) \qquad\qquad
	\notag \\ +\frac{\eta^2 L}{2K^2} {\mathbb E}\left(||\varepsilon^{(t)}||^2\right) + \frac{\eta \rho}{2}||{\mathbb E}\left(\varepsilon^{(t)}\right)||^2 +\frac{\eta^2 L [L (1+\eta) (1-\delta)+ \delta/2]}{\rho \delta \beta^2} G^2 + \frac{\eta^2 L }{2 K^2} \sigma_z^2, \notag \\
	\leq A||{\bf g}^{(t)}||^2  +\frac{\eta^2 L B G^2}{\beta^2 } +\frac{\eta^2 L}{2 K^2} {\mathbb E}\left(||\varepsilon^{(t)}||^2\right) + \frac{C}{K^2}||{\mathbb E}\left(\varepsilon^{(t)}\right)||^2 + \frac{\eta^2 L }{2 K^2} \sigma_z^2,
\end{align}
where
\begin{align}
	&A = \frac{\eta \rho (\eta L^2+ \eta +1)}{2\beta^2} + \frac{\rho}{2} - \frac{(\rho L +1 )\eta}{\beta},\notag \\
	&B = \frac{L (1+\eta) (1-\delta)+ \delta/2}{\rho \delta},\notag \\
	&C = \frac{\eta^2 + \rho^2 \eta +\rho^2}{2\rho} \notag.
\end{align}
With a given value range of hyper-parameters, the following inequality holds:
\begin{align}
	A < -\frac{\eta}{\beta}
\end{align}
Taking average over $t$ on the both sides and rearranging terms yields
\begin{align}
	{\mathbb E}\left[\frac{1}{T}\sum_{t=0}^T ||{\bf g}^{(t)}||^2 \right] 
	\leq \frac{\beta}{\eta}{\left(\frac{F_0-F^*}{ T} + \frac{\eta^2 L B G^2}{\beta^2 } + \frac{\eta^2 L}{2K^2} \sigma_z^2 \qquad\qquad\qquad\qquad\qquad\qquad\quad\right. } \notag \\ + { \left. \frac{\eta^2 L}{2TK^2}\sum_{t=1}^T{\mathbb E}\left[||\bm \varepsilon^{(t)}||^2\right] + \frac{C}{TK^2}\sum_{t=1}^T||{\mathbb E}\left[\bm \varepsilon^{(t)}\right]||^2 \right)},
\end{align}

\subsection{Proof of Lemma \ref{lemma:recurence}} \label{app:lemma 3}
According to the definition of $\hat{\bf w}^{(t)}$, we have
\begin{align} \label{eq:xtp1}
	\hat{\bf w}^{(t+1)}={\bf w}^{(t+1)} - \frac{\eta}{K} \sum_{k=1}^K{{\bf e}_k^{(t+1)}}.
\end{align}
By substituting \eqref{over_the_air_aggre} and  \eqref{eq:new_model_update} to \eqref{eq:xtp1}, we further have
\begin{align}
	\hat{\bf w}^{(t+1)}=&{\bf w}^{(t)}-\eta{\hat y}^{(t)} - \frac{\eta}{K} \sum_{k=1}^K{{\bf e}_k^{(t+1)}}, \notag \\ = &{\bf w}^{(t)}-\frac{\eta}{K} \sum_{k=1}^K{{{ h}}_k^{(t)}  {{ p}}_k^{(t)} {\sf sign}\left({{\bf u}_k^{(t)}}\right)} - \frac{\eta}{K} {\bf z} -\frac{\eta}{K} \sum_{k=1}^K{{\bf e}_k^{(t+1)}}, \notag \\ = &{\bf w}^{(t)}-\frac{\eta}{K} \sum_{k=1}^K{{\tilde{\bf u}}_k^{(t)}} -\frac{\eta}{K} \sum_{k=1}^K{\left({{ h}}_k^{(t)} {{ p}}_k^{(t)}- 1 \right)  {\sf sign}\left({{\bf u}_k^{(t)}}\right)} - \frac{\eta}{K} {\bf z} -\frac{\eta}{K} \sum_{k=1}^K{{\bf e}_k^{(t+1)}}, \notag \\ =& {\bf w}^{(t)}-\frac{\eta}{\beta K} \sum_{k=1}^K{{{\bf g}}_k^{(t)}} - \frac{\eta}{K} {\boldsymbol \varepsilon}^{(t)} - \frac{\eta}{K} {\bf z} -\frac{\eta}{K} \sum_{k=1}^K{{\bf e}_k^{(t)}}, \notag \\ = & \hat{\bf w}^{(t)}-\frac{\eta}{\beta K} \sum_{k=1}^K{{{\bf g}}_k^{(t)}} - \frac{\eta}{K} {\boldsymbol \varepsilon}^{(t)} - \frac{\eta}{K} {\bf z}.
\end{align}
Then we finish the proof.

\subsection{Proof of Lemma \ref{lemma:error_bound}} \label{app:lemma 4}
According to the definition of quantization error in \eqref{accumulated_quantization_error},
\begin{align}
	||{\bf e}_k^{(t+1)} ||^2 = ||{\mathcal C}({\bf u}_{k}^{(t)}) -  {\bf u}_{k}^{(t)}||^2 \leq (1-\delta)||{\bf u}_{k}^{(t)}||^2 = (1-\delta)||\frac{1}{\beta} {\bf g}_{k}^{(t)} +  {\bf e}_{k}^{(t)}||^2.
\end{align}
By using mean-value inequality, we have that:
\begin{align}
	||{\bf e}_k^{(t+1)} ||^2 \leq  (1-\delta)||\frac{1}{\beta} {\bf g}_{k}^{(t)} +  {\bf e}_{k}^{(t)}||^2 \leq (1-\delta)(1+\eta)||{\bf e}_{k}^{(t)}||^2 + \frac{1}{\beta^2} (1-\delta)(1+1/\eta)||{\bf g}_{k}^{(t)}||^2.
\end{align}
Apply the inequality to all communication rounds and we have:
\begin{align}
	{\mathbb E}||{\bf e}_k^{(t+1)} ||^2 & \leq (1-\delta)(1+\eta){\mathbb E}||{\bf e}_{k}^{(t)}||^2 + \frac{1}{\beta^2} (1-\delta)(1+1/\eta){\mathbb E}||{\bf g}_{k}^{(t)}||^2, \notag \\ &\leq \sum_{\tau = 0}^t{\frac{1}{\beta^2} ((1-\delta)(1+\eta))^{t-\tau} (1-\delta)(1+1/\eta) {\mathbb E}||{\bf g}_{k}^{(\tau)}||^2}, \notag \\ &\leq \sum_{\tau = 0}^{\infty}{\frac{1}{\beta^2} ((1-\delta)(1+\eta))^{t-\tau} (1-\delta)(1+1/\eta) G^2}, \notag \\ &= \frac{2(1-\delta)(1+1/\eta)}{\delta \beta^2}G^2.
\end{align}
Now we are ready to proof Lemma \ref{lemma:error_bound}. According to the AM-GM inequality,
\begin{align}
	{\mathbb E}||\frac{1}{K} \sum_{k=1}^K{{\bf e}_k^{(t)}}||^2 &\leq \frac{1}{K} \sum_{k=1}^K{{\mathbb E}||{{\bf e}_k^{(t)}}||^2},
	\notag\\ &\leq \frac{2(1+\eta) (1-\delta)G^2}{\eta \delta \beta^2}.
\end{align}
\subsection{Proof of Corollary 1}\label{app:theo:AWGN}
The proof uses the same framework as Appendix A. The recurrence of $\hat{\bf w}^{(t)} $ in Lemma \ref{lemma:recurence} is rewritten as
\begin{align}
	\hat{\bf w}^{(t+1)}=&{\bf w}^{(t+1)} - \frac{\eta}{K} \sum_{k=1}^K{{\bf e}_k^{(t+1)}} \notag \\
	= &{\bf w}^{(t)}-\eta{\hat {\bf y}}^{(t)} - \frac{\eta}{K} \sum_{k=1}^K{{\bf e}_k^{(t+1)}} \notag \\ = &{\bf w}^{(t)}-\frac{\eta}{K} \sum_{k=1}^K{{\tilde{\bf u}}_k^{(t)}} - \frac{\eta}{K} {\bf z} -\frac{\eta}{K} \sum_{k=1}^K{{\bf e}_k^{(t+1)}} \notag \\ =& {\bf w}^{(t)}-\frac{\eta}{\beta K} \sum_{k=1}^K{{{\bf g}}_k^{(t)}} - \frac{\eta}{K} {\bf z} -\frac{\eta}{K} \sum_{k=1}^K{{\bf e}_k^{(t)}} \notag \\ = & \hat{\bf w}^{(t)}-\frac{\eta}{\beta K} \sum_{k=1}^K{{{\bf g}}_k^{(t)}} - \frac{\eta}{K} {\bf z}
\end{align}
Thus the expected improvement of loss function is given by
\begin{align}
	{\mathbb E}\left[F(\hat{\bf w}^{(t+1)})-F(\hat{\bf w}^{(t)})\right] \leq \frac{\eta(\rho/{2 \beta}-1)}{\beta}||{\bf g}^{(t)}||^2 +  \frac{ 2L (1+\eta) (1-\delta)+\rho\delta}{ 2 \rho \delta \beta^2} \eta^2 L G^2 +\frac{\eta^2 L}{ 2 K^2}\sigma_z^2.
\end{align}
Taking average over $t$ on the both sides and rearranging terms yields
\begin{align}
	{\mathbb E}\left[\frac{1}{T}\sum_{t=1}^T ||{\bf g}^{(t)}||^2 \right]\leq \frac{\beta}{\eta(1-\rho/{2 \beta})}\left(\frac{F({\bf w}_0)-F^*}{T}+\frac{ 2L (1+\eta) (1-\delta)+\rho\delta}{ 2 \rho \delta \beta^2} \eta^2 L G^2 +\frac{\eta^2 L}{ 2 K^2}\sigma_z^2 \right).
\end{align}
Let $\eta = \frac{1}{\sqrt{LT}}$, we have
\begin{align}
	{\mathbb E}\left[\frac{1}{T}\sum_{t=0}^T ||{\bf g}^{(t)}||^2 \right]
	\leq \frac{\beta}{\sqrt{T}(1-\rho/{2 \beta})}\left({\sqrt{L}(F_0-F^*)}+\frac{ DG^2}{{\beta}^2 \sqrt{T}} +\frac{\sqrt{L}}{ 2 \sqrt{T} K^2}\sigma_z^2 \right),
\end{align}
where
\begin{align}
	D = \frac{ 2L (1+1/\sqrt{LT}) (1-\delta)+\rho\delta}{ 2 \rho \delta} \sqrt{L}. \notag
\end{align}

\subsection{Proof of Power Control Optimization}\label{solution}
Given $\boldsymbol {\mathrm P1}$ is a convex problem and the strong duality is hold, it can be solved by Lagrange dual method. The Lagrange function of \eqref{P1} is written as:
\begin{align}
	L(\{p_k\},\{\lambda_k \}) = \frac{(\rho^2 \!+\! \rho^2 \eta \!+\! \eta^2 (\rho L +1))q }{2\rho TK^2} \sum_{t=0}^{T-1}  {\left(\!\sum_{k=1}^K\!\!{h_{k}^{(t)} p_{k}^{(t)}-K\!\!} \right)}^2 \! \qquad \qquad \qquad \qquad \notag \\ + \! \frac{\eta^2 L \| {\boldsymbol{\sigma}_{1}} \|^2 }{2TK^2} \! \sum_{t=0}^{T-1} \! \sum_{k=1}^K \!\!{\left( h_{k}^{(t)} p_{k}^{(t)}-1\right)^2} + \sum_{k=1}^K{\lambda_k \left({|p_k|^2}-\frac{P_0}{M} \right)}.
\end{align}
Then the Lagrange dual function is given by
\begin{align}\label{df}
	g(\{\lambda_k \}) = \underset{\{p_k \geq 0\}}{\inf} L(\{p_k\},\{\lambda_k \}),
\end{align} 
and dual problem is
\begin{align}
	\boldsymbol {\mathrm P2}:&\max \,\,  g(\{\lambda_k \}) \notag \\ &s.t.\,\, \lambda_k \geq 0, \forall k \in {\mathcal{K}} 
\end{align}
By taking the first-order derivative of \eqref{df}, we obtain the optimal solution to $ \boldsymbol {\mathrm P2} $:
\begin{align}
	{p_{k}^*}= \frac{A h_k}{h_k^2 + \frac{2}{\eta^2 L \| {\boldsymbol{\sigma}_{1}} \|^2} \lambda_k^*},,
\end{align}
where 
\begin{align}
	A = \frac{ \rho \eta^2 L \| {\boldsymbol{\sigma}_{1}} \|^2 + \left(\left(\rho L+1\right)\eta^2+\rho^2 \eta +\rho^2\right)q K}{ \eta^2 L \| {\boldsymbol{\sigma}_{1}} \|^2 \left( \rho + {\left(\left(\rho L+1\right)\eta^2+\rho^2 \eta +\rho^2\right)q \sum_{j=1}^K {\frac{h_j}{B_j} }} \right)}
\end{align}
and
\begin{align}
	B_j ={\eta^2 L \| {\boldsymbol{\sigma}_{1}} \|^2 h_{j}} + \frac{2\lambda_j^*}{h_{j}}.
\end{align}
Then we finish the proof.


\begin{thebibliography}{10}
	
	\bibitem{zhu2019broadband}
	G.~Zhu, Y.~Wang, and K.~Huang, ``Broadband analog aggregation for low-latency
	federated edge learning,'' {\em IEEE Transactions on Wireless
		Communications}, vol.~19, no.~1, pp.~491--506, 2019.
	
	\bibitem{park2019wireless}
	J.~Park, S.~Samarakoon, M.~Bennis, and M.~Debbah, ``Wireless network
	intelligence at the edge,'' {\em Proceedings of the IEEE}, vol.~107, no.~11,
	pp.~2204--2239, 2019.
	
	\bibitem{mcmahan2017communication}
	B.~McMahan, E.~Moore, D.~Ramage, S.~Hampson, and B.~A. y~Arcas,
	``Communication-efficient learning of deep networks from decentralized
	data,'' in {\em Artificial intelligence and statistics}, pp.~1273--1282,
	PMLR, 2017.
	
	\bibitem{yang2019federated}
	Q.~Yang, Y.~Liu, T.~Chen, and Y.~Tong, ``Federated machine learning: Concept
	and applications,'' {\em ACM Transactions on Intelligent Systems and
		Technology (TIST)}, vol.~10, no.~2, pp.~1--19, 2019.
	
	\bibitem{samarakoon2019distributed}
	S.~Samarakoon, M.~Bennis, W.~Saad, and M.~Debbah, ``Distributed federated
	learning for ultra-reliable low-latency vehicular communications,'' {\em IEEE
		Transactions on Communications}, vol.~68, no.~2, pp.~1146--1159, 2019.
	
	\bibitem{zhu2021over}
	G.~Zhu, J.~Xu, K.~Huang, and S.~Cui, ``Over-the-air computing for wireless data
	aggregation in massive iot,'' {\em IEEE Wireless Communications}, vol.~28,
	no.~4, pp.~57--65, 2021.
	
	\bibitem{nazer2007computation}
	B.~Nazer and M.~Gastpar, ``Computation over multiple-access channels,'' {\em
		IEEE Transactions on information theory}, vol.~53, no.~10, pp.~3498--3516,
	2007.
	
	\bibitem{soundararajan2012communicating}
	R.~Soundararajan and S.~Vishwanath, ``Communicating linear functions of
	correlated gaussian sources over a mac,'' {\em IEEE Transactions on
		Information Theory}, vol.~58, no.~3, pp.~1853--1860, 2012.
	
	\bibitem{wang2011distortion}
	C.-H. Wang, A.~S. Leong, and S.~Dey, ``Distortion outage minimization and
	diversity order analysis for coherent multiaccess,'' {\em IEEE transactions
		on signal processing}, vol.~59, no.~12, pp.~6144--6159, 2011.
	
	\bibitem{goldenbaum2014channel}
	M.~Goldenbaum and S.~Stanczak, ``On the channel estimation effort for analog
	computation over wireless multiple-access channels,'' {\em IEEE Wireless
		Communications Letters}, vol.~3, no.~3, pp.~261--264, 2014.
	
	\bibitem{goldenbaum2015achievable}
	M.~Goldenbaum, S.~Sta{\'n}czak, and H.~Boche, ``On achievable rates for analog
	computing real-valued functions over the wireless channel,'' in {\em 2015
		IEEE International Conference on Communications (ICC)}, pp.~4036--4041, IEEE,
	2015.
	
	\bibitem{goldenbaum2013robust}
	M.~Goldenbaum and S.~Stanczak, ``Robust analog function computation via
	wireless multiple-access channels,'' {\em IEEE Transactions on
		Communications}, vol.~61, no.~9, pp.~3863--3877, 2013.
	
	\bibitem{abari2015airshare}
	O.~Abari, H.~Rahul, D.~Katabi, and M.~Pant, ``Airshare: Distributed coherent
	transmission made seamless,'' in {\em 2015 IEEE Conference on Computer
		Communications (INFOCOM)}, pp.~1742--1750, IEEE, 2015.
	
	\bibitem{zhu2018mimo}
	G.~Zhu and K.~Huang, ``Mimo over-the-air computation for high-mobility
	multimodal sensing,'' {\em IEEE Internet of Things Journal}, vol.~6, no.~4,
	pp.~6089--6103, 2018.
	
	\bibitem{li2019wirelessly}
	X.~Li, G.~Zhu, Y.~Gong, and K.~Huang, ``Wirelessly powered data aggregation for
	iot via over-the-air function computation: Beamforming and power control,''
	{\em IEEE Transactions on Wireless Communications}, vol.~18, no.~7,
	pp.~3437--3452, 2019.
	
	\bibitem{wen2019reduced}
	D.~Wen, G.~Zhu, and K.~Huang, ``Reduced-dimension design of mimo over-the-air
	computing for data aggregation in clustered iot networks,'' {\em IEEE
		Transactions on Wireless Communications}, vol.~18, no.~11, pp.~5255--5268,
	2019.
	
	\bibitem{sery2020analog}
	T.~Sery and K.~Cohen, ``On analog gradient descent learning over multiple
	access fading channels,'' {\em IEEE Transactions on Signal Processing},
	vol.~68, pp.~2897--2911, 2020.
	
	\bibitem{xu2021learning}
	C.~Xu, S.~Liu, Z.~Yang, Y.~Huang, and K.-K. Wong, ``Learning rate optimization
	for federated learning exploiting over-the-air computation,'' {\em IEEE
		Journal on Selected Areas in Communications}, vol.~39, no.~12,
	pp.~3742--3756, 2021.
	
	\bibitem{sun2021dynamic}
	Y.~Sun, S.~Zhou, Z.~Niu, and D.~G{\"u}nd{\"u}z, ``Dynamic scheduling for
	over-the-air federated edge learning with energy constraints,'' {\em IEEE
		Journal on Selected Areas in Communications}, vol.~40, no.~1, pp.~227--242,
	2021.
	
	\bibitem{xia2021fast}
	S.~Xia, J.~Zhu, Y.~Yang, Y.~Zhou, Y.~Shi, and W.~Chen, ``Fast convergence
	algorithm for analog federated learning,'' in {\em ICC 2021-IEEE
		International Conference on Communications}, pp.~1--6, IEEE, 2021.
	
	\bibitem{fan2021jointa}
	X.~Fan, Y.~Wang, Y.~Huo, and Z.~Tian, ``Joint optimization of communications
	and federated learning over the air,'' {\em IEEE Transactions on Wireless
		Communications}, 2021.
	
	\bibitem{amiri2020machine}
	M.~M. Amiri and D.~G{\"u}nd{\"u}z, ``Machine learning at the wireless edge:
	Distributed stochastic gradient descent over-the-air,'' {\em IEEE
		Transactions on Signal Processing}, vol.~68, pp.~2155--2169, 2020.
	
	\bibitem{amiri2020federated}
	M.~M. Amiri and D.~G{\"u}nd{\"u}z, ``Federated learning over wireless fading
	channels,'' {\em IEEE Transactions on Wireless Communications}, vol.~19,
	no.~5, pp.~3546--3557, 2020.
	
	\bibitem{fan20211}
	X.~Fan, Y.~Wang, Y.~Huo, and Z.~Tian, ``1-bit compressive sensing for efficient
	federated learning over the air,'' {\em arXiv preprint arXiv:2103.16055},
	2021.
	
	\bibitem{cao2021optimized}
	X.~Cao, G.~Zhu, J.~Xu, Z.~Wang, and S.~Cui, ``Optimized power control design
	for over-the-air federated edge learning,'' {\em IEEE Journal on Selected
		Areas in Communications}, vol.~40, no.~1, pp.~342--358, 2021.
	
	\bibitem{yang2022over}
	H.~Yang, P.~Qiu, J.~Liu, and A.~Yener, ``Over-the-air federated learning with
	joint adaptive computation and power control,'' {\em arXiv preprint
		arXiv:2205.05867}, 2022.
	
	\bibitem{zhang2021gradient}
	N.~Zhang and M.~Tao, ``Gradient statistics aware power control for over-the-air
	federated learning,'' {\em IEEE Transactions on Wireless Communications},
	vol.~20, no.~8, pp.~5115--5128, 2021.
	
	\bibitem{guo2022joint}
	W.~Guo, R.~Li, C.~Huang, X.~Qin, K.~Shen, and W.~Zhang, ``Joint device
	selection and power control for wireless federated learning,'' {\em IEEE
		Journal on Selected Areas in Communications}, 2022.
	
	\bibitem{zhu2020one}
	G.~Zhu, Y.~Du, D.~G{\"u}nd{\"u}z, and K.~Huang, ``One-bit over-the-air
	aggregation for communication-efficient federated edge learning: Design and
	convergence analysis,'' {\em IEEE Transactions on Wireless Communications},
	vol.~20, no.~3, pp.~2120--2135, 2020.
	
	\bibitem{karimireddy2019error}
	S.~P. Karimireddy, Q.~Rebjock, S.~Stich, and M.~Jaggi, ``Error feedback fixes
	signsgd and other gradient compression schemes,'' in {\em International
		Conference on Machine Learning}, pp.~3252--3261, PMLR, 2019.
	
	\bibitem{he2016deep}
	K.~He, X.~Zhang, S.~Ren, and J.~Sun, ``Deep residual learning for image
	recognition,'' in {\em Proceedings of the IEEE conference on computer vision
		and pattern recognition}, pp.~770--778, 2016.
   
	
\end{thebibliography}
\end{document}